%% file: main.tex
\documentclass[screen, sigconf]{acmart}
\AtBeginDocument{%
  \providecommand\BibTeX{{%
    \normalfont B\kern-0.5em{\scshape i\kern-0.25em b}\kern-0.8em\TeX}}}

\copyrightyear{2024}
\acmYear{2024}

\setcopyright{cc}
\setcctype[4.0]{by}
\acmConference[CHI '24]{Proceedings of the CHI Conference on Human Factors
in Computing Systems}{May 11--16, 2024}{Honolulu, HI, USA}
\acmBooktitle{Proceedings of the CHI Conference on Human Factors in
Computing Systems (CHI '24), May 11--16, 2024, Honolulu, HI, USA}
\acmDOI{10.1145/3613904.3642730}
\acmISBN{979-8-4007-0330-0/24/05}

\makeatletter

\newcommand\xlabel[2][]{\phantomsection\def\@currentlabelname{#1}\label{#2}}

\makeatother

\usepackage{calc} 
\usepackage{subcaption}
\usepackage{listings}
\usepackage{xcolor}

\newcommand{\rv}[1]{\begingroup\color{black}#1\endgroup}
\newcommand{\bpstart}[1]{\vspace{1mm} \noindent{\textbf{#1.}}}

\begin{document}

\title{MAIDR: Making Statistical Visualizations Accessible with Multimodal Data Representation}

\author{JooYoung Seo*}
\email{jseo1005@illinois.edu}
\affiliation{
  \institution{University of Illinois}
  \city{Urbana-Champaign}
  \state{Illinois}
  \country{USA}
}

\author{Yilin Xia*}
\email{yilinx2@illinois.edu}
\affiliation{
  \institution{University of Illinois}
  \city{Urbana-Champaign}
  \state{Illinois}
  \country{USA}
}

\author{Bongshin Lee}
\email{bongshin@microsoft.com}
\affiliation{
  \institution{Microsoft Research}
  \city{Redmond}
  \state{Washington}
  \country{USA}
}

\author{Sean McCurry}
\email{sean.m.mccurry@gmail.com}
\affiliation{
  \institution{TransPerfect}
  \city{Boulder}
  \state{Colorado}
  \country{USA}
}

\author{Yu Jun Yam}
\email{yujuny2@illinois.edu}
\affiliation{
  \institution{University of Illinois}
  \city{Urbana-Champaign}
  \state{Illinois}
  \country{USA}
}

\thanks{*Two authors contributed equally to this research.}

\renewcommand{\shortauthors}{Seo and Xia, et al.}

\begin{abstract}
\rv{This paper investigates new data exploration experiences that enable blind users to interact with statistical data visualizations---bar plots, heat maps, box plots, and scatter plots---leveraging multimodal data representations.} In addition to sonification and textual descriptions that are commonly employed by existing accessible visualizations, our MAIDR (multimodal access and interactive data representation) system incorporates two additional modalities (braille and review) that offer complementary benefits. It also provides blind users with the autonomy and control to interactively access and understand data visualizations. In a user study involving 11 blind participants, we found the MAIDR system facilitated the accurate interpretation of statistical visualizations. Participants exhibited a range of strategies in combining multiple modalities, influenced by their past interactions and experiences with data visualizations. This work accentuates the overlooked potential of combining refreshable tactile representation with other modalities and elevates the discussion on the importance of user autonomy when designing accessible data visualizations.

\end{abstract}



\maketitle

\input{1-introduction}

\input{2-background}

\input{3-codesign.tex}
\input{4-system}
\input{5-user-study}
\input{6-findings-RQ1}

\input{7-findings-RQ2}
\input{8-discussion}
\input{9-conclusion}
\input{10-acknowledgement}

\bibliographystyle{ACM-Reference-Format}
\bibliography{main}

\input{11-appendix}
\end{document}

%% file: 1-introduction.tex
\section{Introduction}
\label{sec:intro}

The adage ``a picture is worth a thousand words'' reflects the cognitive efficiency conferred by well-designed visual information. As society produces an ever-growing wealth of data, the importance of data visualization is magnified in multiple sectors, ranging from finance and healthcare to politics and education. Yet, data visualization relies on a tacit assumption: the visual ability. This underlying assumption casts a paradox, as what is \emph{convenient for sighted individuals} may morph into \emph{an impediment for blind individuals}\footnote{We use the identity-first language (e.g., blind and low-vision people) instead of the person-first language (e.g., people who are blind; people with vision loss or visual impairments), guided by the perspective of the National Federation of the Blind.}. According to the recent accessibility audit on COVID-19 data visualization revealed that only 14\% of top Google search results visualizations to be highly accessible to screen reader users\footnote{Screen reader users and blind people are interchangeable in this context as blind people require screen readers to access web content.}, with significantly higher pass rates for ``Born Accessible'' websites designed with accessibility in mind \cite{fanAccessibilityDataVisualizations2023}. Similarly, the recent studies by Joyner et al. also highlights the current accessibility challenges in data visualization \cite{joynerVisualizationAccessibilityWild2022}. Over half of the data visualization samples collected from diverse sectors did not notify the existence of the chart or chart type to screen reader users, and the underlying data was inaccessible. In addition, among 144 sighted visualization practitioners, around 38\% said their visualizations are typically not accessible or not very accessible, 23\% said they do not know how accessible their visualizations are.

To address this gap, recent years have witnessed active commitments in HCI, accessibility, and visualization communities. These commitments include conceptual frameworks guiding accessible data visualization \cite{kimAccessibleVisualizationDesign2021}, testable heuristics and audit guidelines for inclusive data visualization \cite{elavskyHowAccessibleMy2022}, and various methods that convert visual information into alternative and non-visual representations \cite{jooyoungseoTeachingVisualAccessibility2023}, such as non-speech sounds (i.e., sonification) \cite{summersAccessibleDataVisualizations2014, julianna-langstonChart2Music2022, HighchartsSonificationStudio2022, summersConvertingGraphicalDatavisualizations2019, hassakuAudioplotlib2022, summersUserInterfacesConverting2022, AudioGraphsApple}, natural language text descriptions \cite{lundgardAccessibleVisualizationNatural2022, aultEvaluationLongDescriptions2002, choiVisualizingNonVisual2019, gouldEffectivePracticesDescription2008}, and tactile graphs \cite{de2021interdependent, brown2012viztouch}.

Particularly, recent work has focused on leveraging multiple representations to complement strengths and weaknesses (e.g., \cite{thompsonChartReaderAccessible2023,sharifVoxLensMakingOnline2022,siuSupportingAccessibleData2022b}). 
While these cross and mixed modalities offer better non-visual accessibility, there have been two limited trends that we want to expand upon. First, recent work on interactive multimodal data visualizations is mostly limited to basic plot types, such as bar plots and time series line plots \cite{thompsonChartReaderAccessible2023, siuSupportingAccessibleData2022b}. This observation holds true even when considering studies focused either on single modality or on static visualizations. According to a recent literature review by Kim et al. \cite{kimAccessibleVisualizationDesign2021}, out of 56 papers published between 1999 and 2020 focused on vision-related accessibility in data visualization, around 80\% of the papers were limited to the four basic plots: bar plots, line plots, pie charts, and point-only scatter plots. Second, the web-based multimodal representation still relies on two conventional methods---sonification and speech-based textual descriptions. This is largely because the web-based visualizations and its multimodal interaction techniques are constrained by the capabilities of current assistive technologies. For example, verbal information, such as simple alt texts, natural language descriptions, lists, and tables, have long been the go-to strategies for accessible non-visual data representation methods over the past decades, facilitated by screen reader's text-to-speech technology \cite{aultEvaluationLongDescriptions2002, gouldEffectivePracticesDescription2008, lundgardAccessibleVisualizationNatural2022}. On the other hand, sonification---the non-verbal information using sound effects---has gained growing interest to complement and overcome the conventional speech-only method \cite{huntInteractiveSonification2011, summersAccessibleDataVisualizations2014}.

To push the boundaries of the current effort of multimodal and accessible data visualization, we introduce a novel approach representing data visualization patterns and shapes in refreshable braille displays (RBDs), which can be seamlessly integrated into blind users' existing computer access workflows as part of multimodal data representation methods. Although some researchers pioneered static 3d-printed or embossed tactile graphics overlay to touchscreen device to produce audio-tactile graphics \cite{gotzelmannVisuallyAugmentedAudioTactile2018, heTacTILEPreliminaryToolchain2017}, no study has yet addressed refreshable, interactive, and automatic tactile representation methods using blind users' computer access technology in the context of statistical visualizations.

\rv{Informed and inspired by everyday challenges a blind data science educator experienced} in various statistical visualizations at work, our mixed-ability team has co-designed and co-developed the Multimodal Access and Interactive Data Representation (MAIDR) system that combines auditory and tactile representation verbally and non-verbally with data visualization. By employing four modes---Braille, Text, Sonification, and Review (BTS+R)---MAIDR aims to offer a holistic framework for data interaction that accommodates a spectrum of visual abilities. 
In this paper, we introduce how MAIDR facilitates the interaction with four types of statistical plots---bar plots, heat maps, box plots, and smooth-line-layered scatter plots\rv{\footnote{In this paper, smooth-line-layered plots refer to scattered point plots with a smoothed prediction line overlaid.}}---each of which is tied to its own unique statistical concepts.

We conducted a user study with 11 blind participants with an aim to answer the following two research questions: 
(RQ1) What strategies do blind users employ to combine multiple modalities (BTS+R) in their data exploration processes within and across four statistical visualizations? and (RQ2) To what extent, can RBDs be effectively incorporated into the existing computer access workflows for blind users in the context of data visualizations?
We asked our blind participants to address four common types of data interpretation (i.e., lookup, compositional, visual, and non-visual) questions \cite{kimAnsweringQuestionsCharts2020} in the four statistical visualizations. 
Our qualitative and quantitative results reveal that the MAIDR system facilitated the accurate interpretation of statistical visualizations and participants exhibited a range of strategies in combining multiple modalities, influenced by their past interactions and experiences with data visualizations. This work accentuates the overlooked potential of combining refreshable tactile representation with other modalities and elevates the discussion on the importance of user autonomy when designing accessible data visualizations.

%% file: 2-background.tex
\section{Background and Related Work}
\label{sec:background}

\subsection{Theoretical Background}
\label{sec:theories}

Our work is situated at the intersection of two foundational theoretical frameworks: Multimodality \cite{kressMultimodalitySocialSemiotic2010} and the Cognitive Theory of Multimedia Learning (CTML) \cite{mayerCognitiveTheoryMultimedia2014}. Multimodality posits that human communication encompasses a rich tapestry of semiotic channels, or ``modes.'' These modes are culturally and socially recognized avenues for expression and can include written text, gestures, postures, and gazes, as well as visual elements like font and color \cite{kressMultimodalitySocialSemiotic2010}.

Parallel to this, advances in web-based multimedia have stimulated scholarly exploration into the cognitive implications of multimodal communication. Mayer's CTML \cite{mayerCognitiveTheoryMultimedia2014} outlines instructional design principles optimized for cognitive load management in multimedia learning environments. The theory is undergirded by three key assumptions: (a) the Dual-Channel Assumption, stipulating separate channels for processing auditory-verbal and visual- pictorial information \cite{clarkDualCodingTheory1991}; (b) the Limited-Capacity Assumption, suggesting that each channel can only process a finite amount of information \cite{swellerCognitiveLoadProblem1988}; and (c) the Active Processing Assumption, emphasizing that effective learning requires the orchestration of cognitive processes like selection, organization, and integration of information across modalities \cite{swellerElementInteractivityIntrinsic2010}.


\rv{
Data science is a multidisciplinary field that integrates various modes of communication. However, visual methods such as graphs, tables, colors, and fonts remain predominant in data science teaching and learning, similar to other STEM subjects. This preference is due to the cognitive benefits these visual aids provide in human learning, as described in CTML \cite{mayerCognitiveTheoryMultimedia2014}. Nevertheless, we argue that this CTML framework, which is based on a dual coding theory involving pictorial and auditory elements, should be carefully redesigned for blind people. Without such redesign, blind learners might be overly dependent on verbal information alone.
}

\subsection{Alternative Data Visualization Methods for Blind People}
\label{sec:review}

In this section, we will review existing literature on multimodal data visualization strategies tailored for blind people and delineate how our research contributes novel perspectives to this discourse.

\subsubsection{Textual Descriptions}
\label{sec:textual_description}

In the data visualization context, visual components (e.g., colors, shapes, size) are processed through the non-verbal channel while text-based ones (e.g., data values, labels) are processed through the verbal channel. For blind people, textual descriptions using verbal channel have \rv{long been suggested to access data visualization due to its simplicity and compatibility with screen readers.

In 2002, for instance, Ault et al. evaluated the effectiveness of textual descriptions for conveying common graphs like bar, line, and pie charts to blind learners on the web and proposed guidelines to optimize accessibility and comprehension of statistical graphs and tables \cite{aultEvaluationLongDescriptions2002}. Six years  later, a formal guidelines by the National Center for Accessible Media (NCAM) suggested describing STEM images and data charts in list items and tables \cite{ncam2008}.} Using the NCAM guidelines, Morash et al. developed and evaluated a template-based description generator for data charts which lead to more standardized word usage and structure \cite{morash2015guiding}. More recently, \rv{Jung et al. systematically revisited strategies for formulating effective alternative text (alt text) for data visualizations to enhance accessibility for blind and low-vision people by investigating misalignment between guidelines and practice \cite{jungCommunicatingVisualizationsVisuals2022}.

In the domain of statistical computing, Godfrey et al. introduced the BrailleR package offering a programmatic text description method for blind R users to access base R and ggplot2 graphics \cite{braille2018}. Additionally, they extended the package to include an accessible interaction text-tree model \cite{godfrey2018accessible}.

On the other hand, opportunities have been also suggested to extend the textual description using linguistics and general structure. For example,} Lundgard and Satyanarayan \cite{lundgardAccessibleVisualizationNatural2022} introduced a four-level model for categorizing the semantic content conveyed by natural language descriptions of visualizations---(L1) basic descriptions of a visualization's construction, (L2) reporting statistical facts and relationships, (L3) identifying perceptual trends and phenomena, and (L4) providing contextual insights and interpretations. \rv{Zong et al. further developed these natural language descriptions into three key dimensions on rich screen reader experiences (i.e., structure, navigation, and description) \cite{zongRichScreenReader2022a}. This led to the development of Olli, a JavaScript library that converts existing data visualizations into keyboard-navigable treeview structures with text descriptions to make them accessible to screen reader users \cite{blancoOlliExtensibleVisualization2022}. While not exclusively focused on textual description alone, a node-link graph structure, called Data Navigator, was also proposed to enable visualization authors to generically represent diverse data relationships, such as trees, flows, and geographic maps in assistive technology compatible forms \cite{elavskyDataNavigatorAccessibilityCentered2023}.

Although a textual description using screen reader is an effective method to access data visualization, relying on the verbal channel alone might be cognitively overloading when it gets longer and nested. For instance, a recent study found that blind people were more efficient in answering comprehension questions requiring understanding sonification patterns than textual descriptions~\cite{siuSupportingAccessibleData2022b}. Therefore, we propose to combine textual description with sonification to support blind people's data visualization access.
}

\subsubsection{Tactile Graphics}
\label{sec:tactile_graphics}

Tactile graphics have been a non-verbal conventional method used by blind people to access data visualization along with textual description. Although it is an effective alternative to visualization, tactile graphic productions often require sighted tactile designers' help to properly format ready-to-print files compatible with braille embosser, swell form heating machine, or 3D printer \cite{de2021interdependent}.

While traditional embossed tactile graphs lack multimodal interactivity, recent work has investigated this by using audio-tactile approach. For example, He et al. \cite{heTacTILEPreliminaryToolchain2017} introduced TacTILE, a toolchain to create tactile overlays with audio annotations for touchscreen graphics to improve accessibility for blind users. 
In a similar vein, Götzelmann \cite{gotzelmannVisuallyAugmentedAudioTactile2018} presented an approach for creating visually augmented audio-tactile graphics to support blind individuals. 

Although these novel audio-tactile approaches show promise in improving accessibility of graphical information for blind individuals through multimodal interaction, it  still requires manual production of tactile graphs which requires a sighted assistance and the static overlay cannot be easily integrated into dynamic visualizations on the web.

\rv{
To address this issue, we propose the use of refreshable braille displays (RBDs) to support tactile graphics. Over the last few decades, RBDs have become a pivotal assistive technology along with screen reader for blind individuals having braille literacy, offering efficient access to digital content \cite{lu2015,kapperman2021}. Unlike braille embossers and swell form machines, RBDs provide real-time access to information, enabling users to read and write braille dynamically \cite{lu2015}. Coupled with a screen reader, RBDs have enhanced educational and employment opportunities \cite{kapperman2021,whiteAccessibilityMathematicalNotation2020,cervoneAdaptableAccessibilityFeatures2019}. 

While RBDs have been extensively used for text-based verbal information, research has shown the potential of RBDs to represent graphical information, including graphs and charts, through multi-line braille displays \cite{frediani2018}. For instance, the use of electroactive elastomers in portable multiple-line RBDs has been explored, highlighting the potential for enhancing the accessibility and portability of Braille materials, including those related to mathematics \cite{frediani_enabling_2018}. Orbit Research has also released refreshable tactile graphics display, called Graphiti \cite{GraphitiBreakthroughNonVisual} with 60 by 40 grid of tactile pins. Furthermore, the effectiveness of non-verbal braille displays has been verified through experimental results, demonstrating their potential for supporting mobile use and providing tactile feedback for blind individuals \cite{sawada_displaying_2012}. Additionally, the utilization of shape memory alloy (SMA) wires for braille displays has been investigated, with experiments confirming the effectiveness of this non-verbal approach for presenting Braille characters \cite{zhao_novel_2012}.

However, multi-line RBDs has not been widely adopted among blind people compared to single-line displays \cite{lu2015,frediani_enabling_2018}. This may be attributed to factors such as cost, size, and complexity, which could affect the portability and usability of multi-line displays. Additionally, the limited availability of content optimized for multi-line displays may also contribute to their lower adoption rates.
On the other hand, attempts have been made to represent data visualizations using Unicode braille dot patterns \cite{BraillePatterns,gargBraille8UnifiedBraille2016} and display them on more widely adopted single-line RBDs. For example, the Spark Braille project \cite{SparkBrailleBrailleLine} converted line charts into Unicode braille dot patterns and implemented braille cursor routing interaction. However, it was limited to line charts and did not include other modalities. The Accessible Graph Project (https://accessiblegraphs.org) developed a web-based tool to convert bar plots into braille dot patterns, combining them with sonification through an interactive web interface. However, it was limited to one-vector numeric variables and did not allow for selective toggling of the braille patterns without sonification. Our work expands on these efforts by supporting more complex graphs, such as heat maps, box plots, and scatter plots on single-line RBDs. We also consider the adjustable braille column size on our system to accommodate different types of braille displays.
}

\subsubsection{Sonification and Mixed Modalities}
\label{sec:sonification_review}

Sonification has emerged as an alternative non-verbal data visualization method over the past decades. The Sonification Handbook incorporated past and recent work on multiple ways of mapping data points to sound axes, such as spatial audio and different pitches and sound effects in relation to human cognition \cite{SonificationHandbookEdited}. In data science and statistical visualization context more specifically, SAS Graphic Accelerator has been one of the most well-known sonification tools among blind  people \cite{SASHelpCenter}. It was introduced as a Chrome browser plug-in where users can import their local data file or parse HTML data table to create accessible sonification graphs \cite{summersConvertingGraphicalDatavisualizations2019,summersUserInterfacesConverting2022,summersAccessibleDataVisualizations2014}. The SAS Graphic Accelerator used stereo-audio panning to map horizontal values on x axis and different pitches to  represent vertical values on y axis for cartesian plots, such as bar plots, line plots, scatter plots. It also introduced an innovative way of representing an navigating geospatial data by utilizing a tangible thumbstick as a 360-degree sweepable ``virtual cane'' controller and mapping distance sound with different tones \cite{summersUserInterfacesConverting2022}. 
The SAS Graphic Accelerator has played a great role model for subsequent commercial and open-source data sonification tools, including HighCharts \cite{HighchartsSonificationStudio2022}, Apple's Audio Graphs \cite{AudioGraphsApple}, Chart2Music \cite{julianna-langstonChart2Music2022}, and audio-plot-lib \cite{hassakuAudioplotlib2022}.

\rv{Combining both verbal (textual descriptions) and non-verbal (sonification) modalities, recent work has focused on multimodal benefits. For instance, Siu et al. investigate audio data narratives as a method to improve data communication for screen reader users \cite{siuSupportingAccessibleData2022b}.
Their study results indicate the narrative representation helps participants gain a more complete understanding of the data by integrating both description and sonification. Participants are also more efficient in answering comprehension questions requiring understanding sonification patterns.
Simiarly, Thompson et al. presented Chart Reader \cite{thompsonChartReaderAccessible2023}, a web-based accessibility engine that enables screen reader users to interactively read and understand data visualizations through textual descriptions, sonification, and flexible navigation.
Sharif et al. developed VoxLens, an interactive JavaScript Chrome plug-in that utilizes voice recognition and sonification to enhance the exploration of data visualizations for screen reader users \cite{sharifVoxLensMakingOnline2022}.
The study found that with VoxLens, screen reader users improved their accuracy of extracting information from data visualizations by 122\% and reduced their overall interaction time by 36\% compared to not having the tool. 
}

Taken together, while recent research has been increasingly adopting multimodal data representation methods, verbal textual description and data narration and non-verbal sonification have been a primary combination. Our work introduces a novel approach combining non-verbal tactile representation using RBDs to support statistical data visualizations. 
Unlike sighted people who can utilize multiple senses and strategies to interpret different types of data representations, blind people are often limited to one or two senses of delivery, which calls for the need for an integrated approach. Our proof-of-concept system makes this integration possible by providing learners of varying degrees of disabilities with customizable multimodalities. 

%% file: 3-codesign.tex
\section{Co-Design Process and Considerations}
\label{sec:codesign}

The multimodal access and interactive data representation (MAIDR) system emerged from a  co-design process, leveraging the expertise of a mixed-ability team. This design process was grounded in the interdependence framework \cite{bennett2018interdependence}, which views disabilities as a collective resource. This perspective fosters collaboration between individuals with and without disabilities, channeling their unique experiences and perspectives into the creation of accessible technologies. This method of interdependent co-design mirrors practices in prior works in accessibility field, such as \cite{srinivasanAzimuthDesigningAccessible2023,leeCollabAllyAccessibleCollaboration2022}.

\subsection{Interdependent Design Team}
\label{sec:interdependent_design}

Our interdependent team consisted of three researchers—one blind (BR) and two sighted (SR1 and SR2)— along with two sighted developers (D1 and D2). BR is an early-career researcher and educator in the data science field. He has utilized various screen readers and refreshable braille display products for over 23 years. He had a continual need to access data visualizations in both research and teaching. To meet this need, BR adopted various non-visual strategies and modalities, such as (a) listening to data patterns through sonification tools (e.g., SAS Graphic Accelerator; an R package called `sonify'), (b) touching tactile representations produced with a braille embosser and swell form heating machine, (c) relying on verbal descriptions from human readers, and (d) reading static alternatives like tables and textual summaries. 
 Although each non-visual method had its own strengths, BR found that a system allowing connected and interactive multimodalities was still missing, which would be more effective in accessing data visualizations. This gap inspired our team to embark on the project initially inspired by BR's personal experience but eventually expanding 
 to address broader challenges faced by the blind community in academic and professional settings regarding data visualizations.


SR1 and SR2 brought expertise in accessibility, data visualization, and human-computer interaction (HCI). They played key roles in defining research questions, refining the system design, and formulating study designs. D1 and D2 focused on the technical implementation, ensuring that the system adhered to essential accessibility guidelines.

\subsection{Design Process}
\label{sec:design_process}

The co-design process began in June 2022 with brainstorming sessions and conceptual design discussions, and continued into the development of an early system prototype in August to December 2022. We refined our design goals and considerations while engaging in rapid iterative prototyping and evaluation \cite{medlockUsingRITEMethod2007} with other blind users from January to May 2023 (Section~\ref{sec:study}). To preserve the project’s multidisciplinary nature, our team held two types of meetings: (1) development-focused gatherings to address technical specifics, which happened twice a week and (2) weekly sessions centered on design and evaluation of the system. During research-focused meetings, we critically analyzed prior work on accessible data visualization \cite{sharifVoxLensMakingOnline2022,HighchartsSonificationStudio2022,thompsonChartReaderAccessible2023,SASHelpCenter,gotzelmannVisuallyAugmentedAudioTactile2018,heTacTILEPreliminaryToolchain2017}, assessing strengths and limitations to identify where our project could make meaningful contributions. Each meeting typically lasted about one hour on Zoom. Our team created a shared Google document to record meeting notes and action items, which was updated after each meeting. Based on the outcome from the research meeting, development-oriented meeting discussed the technical details of potential features or improvements. GitHub repository was used to manage the project's codebase, and we used GitHub Issues and pull requests to track bugs and feature requests. BR participated in all meetings, facilitating communication between the subgroups. As a regular user of screen readers and RBDs, his early detection of potential accessibility challenges enabled the team to proactively address them, ensuring broader compatibility with various assistive technologies and web browsers.

\subsection{Design Considerations}
\label{sec:design_consideration}

Through our iterative co-design sessions and literature review, we identified six design considerations (DC) for accessible data visualization system. These considerations guided the development of our MAIDR system, which is described in the next section.

\bpstart{DC1. The system should initially support widely recognized data visualizations, representing distinct statistical concepts} In our initial design phase, we focused on four visualizations: bar plots, heat maps, box plots, and smooth-line-layered scatter plots because of their widespread use in illustrating various statistical concepts\footnote{Post-study, we integrated histograms, line plots, and segmented bar plots into MAIDR and plan to continue expanding our range of visualizations.}. For instance, bar plots, often used to depict frequency distributions, require one categorical and one numerical variable. They are vital for understanding frequency tables and descriptive statistics. Heat maps visualize the relationship between two categorical variables and aid in understanding chi-square concepts. Box plots provide an in-depth view of the descriptive statistics of a numerical variable across categories, relating to concepts like interquartile range and outliers. Scatter plots, requiring at least two numerical variables, are instrumental in exploring correlation and regression analyses. 

In selecting these visualizations, we aimed to minimize overlap in their design insights. For this reason, we initially excluded histograms and line plots from our user studies. Histograms were excluded due to their similar interaction potentially yielding comparable design insights as bar plots. Similarly, the  smoothed prediction line in our scatter plots could replicate the function of a line plot.
We recognize that our selection is not all-encompassing. Other visualizations can convey similar or enhanced statistical concepts. For example, mosaic plots could substitute heat maps for analyzing two categorical variables, violin plots could offer a more detailed view than box plots for numerical data points' interquartile range and density, and hex plots could provide a richer understanding of the relationship between two numerical variables than scatter plots. However, our choices were guided by simplicity and our goal to cover basic and commonly used statistical visualizations that blind individuals frequently encounter.

\bpstart{DC2. Scalable and adaptive tactile representations should be provided, compatible with single-line refreshable Braille displays} To take full advantage of available assistive technologies that are commonly encountered by blind people, we incorporated screen readers and single-line refreshable braille displays (RBDs).
As highlighted in Section~\ref{sec:tactile_graphics}, single-line RBDs, along with screen readers, have become crucial assistive technologies in the education and employment sectors for blind people \cite{lu2015,kapperman2021}. 
We thus sought to develop a non-verbal braille representation that could easily scale up to accommodate newer multi-line braille displays. Most RBDs feature an eight-dot braille chord, including cursor keys, which is also supported by modern unicode braille \cite{BraillePatterns,gargBraille8UnifiedBraille2016}. Therefore, we chose to use eight-dot braille in our design to enable the tactile representation of a wide range of visual patterns within the constraints of conventional braille's low resolution, which aligns with prior work \cite{SparkBrailleBrailleLine}. Furthermore, While six-dot braille supports 64 combinations, eight-dot braille allows for 256 different shapes, a fourfold increase. 

\bpstart{DC3. Each data point should be navigable by standard four-way arrow keys} Interactive data navigation is crucial as it enables users to explore data, uncovering patterns and insights. For instance, users can investigate relationships between variables, identify outliers, or examine closely grouped data points. This level of exploration is not feasible with static alternatives like audio files, tables, or textual summaries. Building on previous work, such as the SAS Graphic Accelerator~\cite{SASHelpCenter}, we used the arrow keys for visualization navigation.

\begin{figure*}[!ht]
  \centering
  \includegraphics[width=0.85\linewidth]{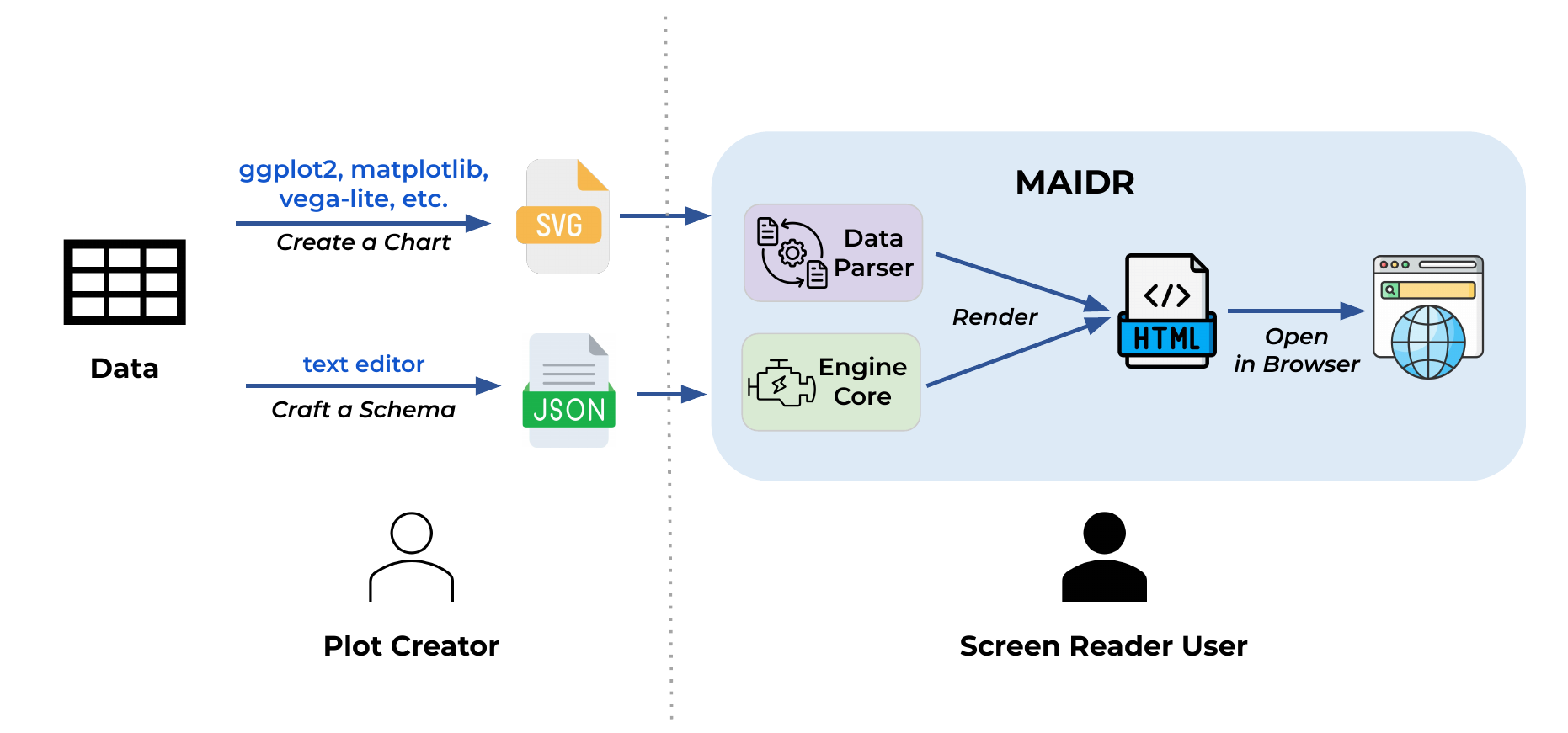}
  \caption{With the chart and schema files as the input, the MAIDR system automatically generates an HTML file containing an accessible visualization, that can be readily rendered and opened in common web browsers.}
  \Description{The image is a flowchart that illustrates the process of converting data into an accessible visualization with MAIDR. It starts with 'Data' on the left, which goes into 'SVG Creation' through tools like 'ggplot2, matplotlib, vegalite, etc.', resulting in an 'SVG' file. Below this process, a note indicates that 'JSON' requires creators to 'Manually Edit'. This JSON then feeds into 'MAIDR' data parser and engine core, which auto-generates 'HTML'. Finally, the HTML is represented as being rendered and ready to 'Open in Browser', symbolized by a web browser icon. The flow is from left to right, with arrows indicating the direction of the process.}
  \label{fig:maidr-system-diagram}
\end{figure*}

\bpstart{DC4. The combination of mixed modalities should be customizable, with a tethered cursor position across modalities}
Previous research on accessible data visualizations has typically concentrated on single modalities or fixed combinations thereof. We aimed to explore how varying combinations of modalities could create unique experiences for blind users. For instance, some might prefer a blend of non-verbal sonification with verbal speech output, while others may choose 
We wanted to provide different granularity levels of information to users with different modalities. For example, users can get a quick overview of the trends of data points using non-verbal tactile modality, and hear the nuanced increase and decrease of data points using non-verbal sonification. Then, users may get more detailed information about the data points using verbal speech output. We wanted to observe how our users would use different modalities and their combinations to explore the data with this different granularity with our system. 
In addition, to allow users to seamlessly switch between modalities without losing track of their current data point, we prioritized a unified system cursor position across all modalities, rather than separate cursors for each modality.

\bpstart{DC5. Users should be allowed to adjust information presented in each modality}
Our co-design process and review of previous works highlighted the importance of allowing users to adjust information in each modality. For instance, since the column size of refreshable braille displays varies among blind users, our system had to be designed to adapt tactile data patterns to fit the users' RBDs size. In textual data representation, users should have the option to choose the verbosity level. This could range from displaying only the current data value to including additional information like x- and y-axis labels. In addition, the sonification aspect should offer customization in terms of pitch and volume, allowing users to tailor it to their auditory preferences.

\bpstart{DC6. Layer-by-layer navigation should be enabled for sequentially accessing overlaid elements
} In layered plots, while sighted users can easily discern multiple geometrical elements overlaid, such as a smoothed line over scatter points, blind users may find navigating non-visually these combined elements overwhelming due to the density of information. To mitigate cognitive overload and potential confusion, we decided to facilitate the ability for blind users to navigate through layers sequentially. 
This method is influenced by the grammar of graphics concept \cite{wickhamLayeredGrammarGraphics2010,wilkinsonGrammarGraphics2012}, which emphasizes the stratification of geometrical layers.

%% file: 4-system.tex
\section{The MAIDR System}
\label{sec:system}

\rv{
MAIDR is a JavaScript library designed to offer a versatile range of multimodal and interactive controls for statistical graphs in web browsers. In its preliminary implementation, it features compatibility with four major types of statistical visualizations: bar plots, heat maps, box plots, and smooth-line-layered scatter plots (\textbf{DC1}).
In this section, we start with a concise overview of the input files needed by MAIDR to provide accessible visualizations. We then explain how blind users can engage with and navigate through the visualizations. Following this, we detail each modality, illustrating how end users can analyze and understand patterns.

\subsection{User Interaction and Navigation}
\label{sec:system_overview}

The MAIDR system expects two input files---the plot image and its corresponding metadata schema---to create an accessible visualization that can be consumed in modern web browsers (Figure~\ref{fig:maidr-system-diagram}).
People can first generate and save a plot as an image file using their preferred visualization package, such as R's ggplot2, Python's matplotlib, Vega-lite, among others, then craft a schema document that describe the plot (in the JSON format in accordance with the template outlined in Appendix~\ref{sec:appendix_json}). The JSON schema, inspired by Chart2Music~\cite{julianna-langstonChart2Music2022} and Vega-lite~\cite{satyanarayan2016vega}, details necessary properties, such as plot type, plot title, axes' labels and legend, and each data point. The plot image and JSON schema are assigned a unique identifier (i.e., id property), which are used to link the two files together in an HTML rendering process. MAIDR is designed to be visual-agnostic, supporting both raster (e.g., PNG) and scalable vector graphics (SVG) objects. However, the visual highlight sync feature is only available for SVG objects, and Target data points must be defined in the JSON schema using CSS selectors.
With these two files the MAIDR system automatically generates an HTML file containing an accessible visualization that can be readily rendered and opened in common web browsers, such as Firefox, Google Chrome, Microsoft Edge, and Safari. This facilitates the effortless sharing of the accessible visualization with the intended audience, thereby allowing them to interact with it.}


\begin{figure*}[!ht]
  \centering
  \includegraphics[width=0.85\linewidth]{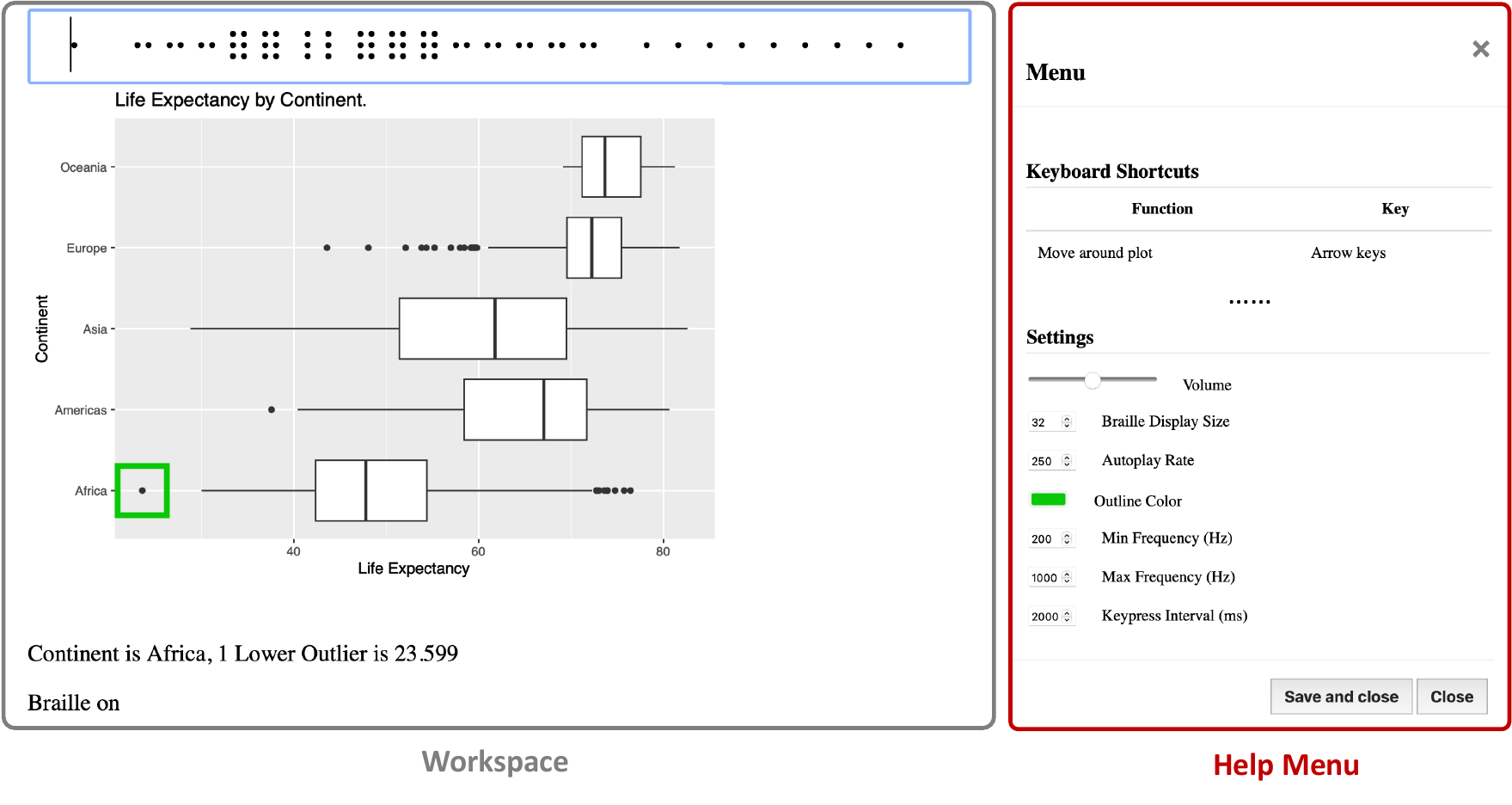}
  \caption{The MAIDR interface, highlighting workspace and help menu components.}
  \Description{This figure illustrates the MAIDR-generated HTML interface. On the left, there is the workspace, the primary interaction zone, featuring a braille pattern section, a visualization area, and a text area for descriptions and modality cues. On the right, there is a help menu detailing keyboard shortcuts for the MAIDR system and specific settings, including the number of braille characters. }
  \label{fig:interface}
\end{figure*}

Upon entering the MAIDR-generated web page, blind users can begin their interactive experience by pressing the `Tab' key to focus on the workspace, the central area designated for visualizations (see Figure~\ref{fig:interface}). \rv{
Building on previous work, such as the SAS Graphic Accelerator~\cite{SASHelpCenter} and our design consideration (\textbf{DC3}), we assigned the left and right arrow keys for navigating data points along the x-axis and the up and down keys for the y-axis. This was straightforward for single-point-per-coordinate visualizations like bar plots and heat maps. However, it was more complex for multi-point visualizations like box plot outliers and scatter plot points. For example, in the box plot, we designed four-way navigation for the interquartile range and whiskers using a fixed grid template. Users navigate various statistical values (e.g., minimum, Q1, median, Q3, maximum) using left or right arrows, and switch between groups for comparison using up or down arrows. For outliers in box plots and y-axis values in scatter plots, accessing them as separate data points was challenging without normalizing the data, which could introduce interpretation bias. Thus, we grouped multiple data points within their corresponding categories. In box plots, outliers below the minimum were labeled as ``lower outliers," and those above the maximum as ``upper outliers." In scatter plots, we grouped y-axis data points sharing the same x-axis value. For instance, points like (1, 2.1), (1, 3.5), (1, 4.0) were grouped as (1, [2.1, 3.5, 4.0]), allowing navigation through data points using the left and right arrows.
}
\rv{In the case of smooth-line-layered scatter plots, the PageUp/PageDown keys enable layer-switching, reflecting \textbf{DC6}. A complete set of advanced navigation shortcuts is documented in the Appendix~\ref{sec:appendix_shortcuts}.
}

The Help menu, which can be invoked by the 'H' key, serves as an all-encompassing guide, offering not only functional key references for quick recall but also a suite of customization settings. These settings range from volume control and braille display size adjustments to the selection of preferred sonification frequency ranges \rv{(\textbf{DC5})}.

By default, all four modalities—Braille, Text, Sonification, and Review—are turned off. This design decision was made to capture user preferences, requiring them to manually activate their desired modalities using specific hotkeys (B, T, S, R). Once a modality is activated, its status is audibly communicated to keep users informed. Subsequent sections provide an in-depth examination of these modalities and their applications across various statistical visualizations.

\subsection{BTS+R Modalities}
\label{sec:btsr_modalities}

Drawing from \rv{theoretical background \cite{kressMultimodalitySocialSemiotic2010,mayerCognitiveTheoryMultimedia2014} and our design consideration (\textbf{DC4}),}
 the MAIDR system is designed to offer four specialized data representation modalities, collectively termed BTS+R, to enable blind users to navigate and engage with statistical visualizations autonomously. This design approach aims to optimize cognitive load by effectively utilizing both verbal and non-verbal channels, which include auditory and tactile representations, as detailed in Table~\ref{table:btsr}.

The primary modalities—Braille, Text, and Sonification—can be toggled on or off with each single-letter key, allowing for up to 8 \(2^3\) unique combinations. These primary modalities also share a synchronized cursor position, enabling blind users to feel (B), read (T), and hear (S) the currently focused data point simultaneously. Review, denoted with `+' symbol, serves as an auxiliary mode (Section~\ref{sec:review_mode} for more details).

\renewcommand\arraystretch{1.3}
\begin{table}[ht]
  \caption{MAIDR System's BTS+R Modalities for Blind Users. The modalities marked with * are primary, while the modality marked with + is auxiliary.}
  \label{table:btsr}
  \footnotesize
  \begin{tabular}{p{1cm}p{3cm}p{3cm}}
    \toprule
                           & Verbal Channel                                                    & Non-Verbal Channel                                                                                \\
    \midrule
    Tactile Representation & R+: Review (literary braille for data values and labels)          & B*: braille (haptic dot patterns representing graphical shapes and trends)                        \\
    \hline
    Sound Representation   & T*: Text  (spoken data values and labels via speech synthesizers) & S*: Sonification  (non-speech, spatial sounds for illustrating data patterns, shapes, and trends) \\
    \bottomrule
  \end{tabular}
\end{table}

\subsubsection{Braille}

The Braille modality in MAIDR serves as a non-verbal channel, offering tactile interpretations of various shapes and patterns present in statistical visualizations \rv{(\textbf{DC2})}. Leveraging specialized algorithms for data-to-dot conversion, the system encodes these visual forms into Unicode Braille characters \cite{gargBraille8UnifiedBraille2016}.  Users can physically feel these tactile representations via their refreshable Braille displays. 
\rv{
  Recognizing that refreshable braille displays vary in cell size (width) across different products, we enabled users to select the appropriate column size for their device in the MAIDR system help menu (\textbf{DC5}). If a braille pattern exceeds the chosen column size, it automatically wraps to the next line, allowing users to navigate between patterns using their braille display's scroll key. This approach guarantees that our system is seamlessly adaptable for use with braille displays of any size, encompassing various numbers of cells and lines, providing users who have access to more braille cells and lines the flexibility to adjust the granularity of the braille patterns they interact with.
}

\begin{figure*}[ht]
  \centering
  \subcaptionbox{\textcolor{black}{Bar plot}\label{fig:braille_verbose_a}}[.49\linewidth][c]{%
    \includegraphics[width=.85\linewidth]{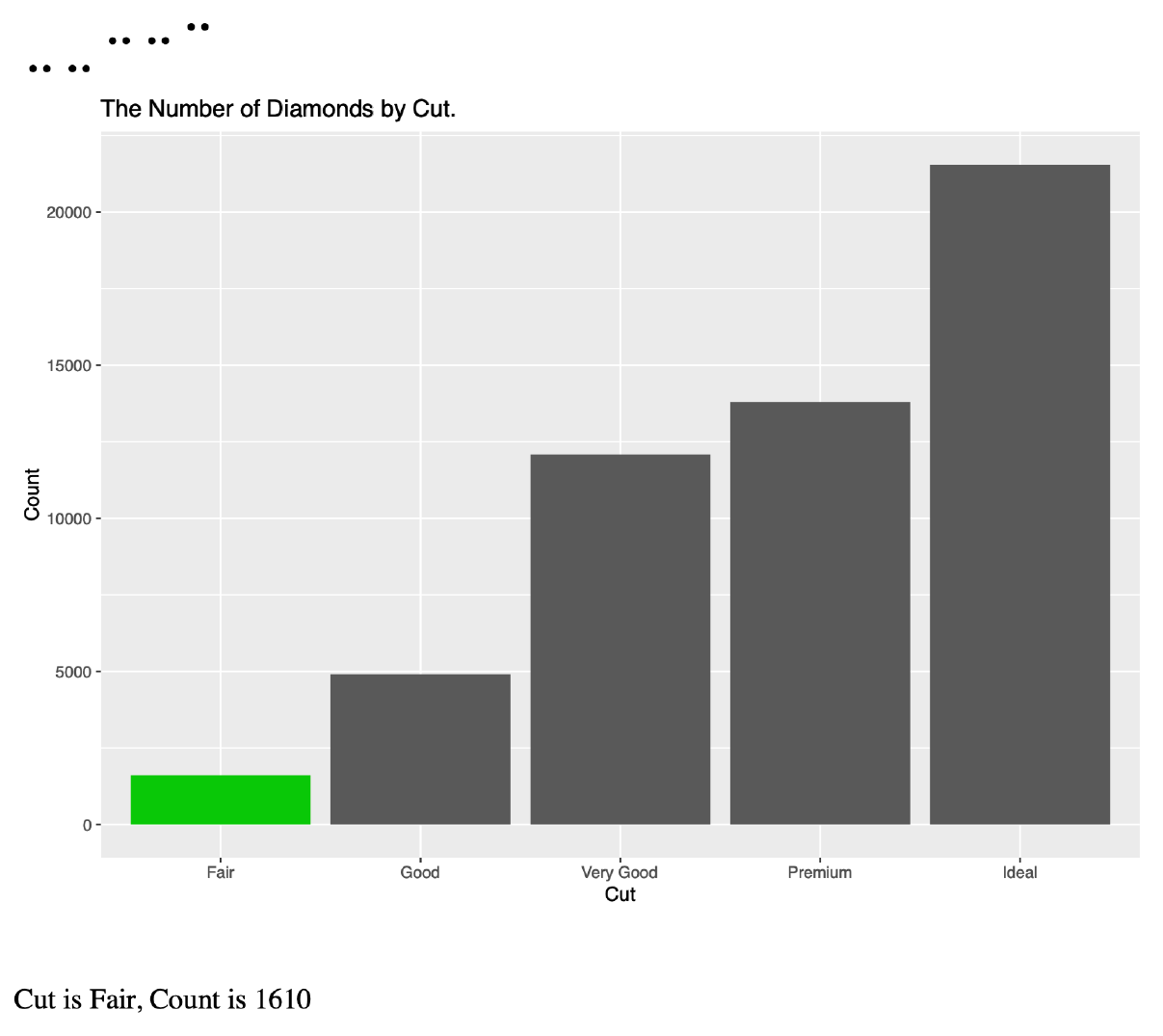}
  }\hfill
  \subcaptionbox{\textcolor{black}{Heat map}\label{fig:braille_verbose_b}}[.49\linewidth][c]{%
    \includegraphics[width=.9\linewidth]{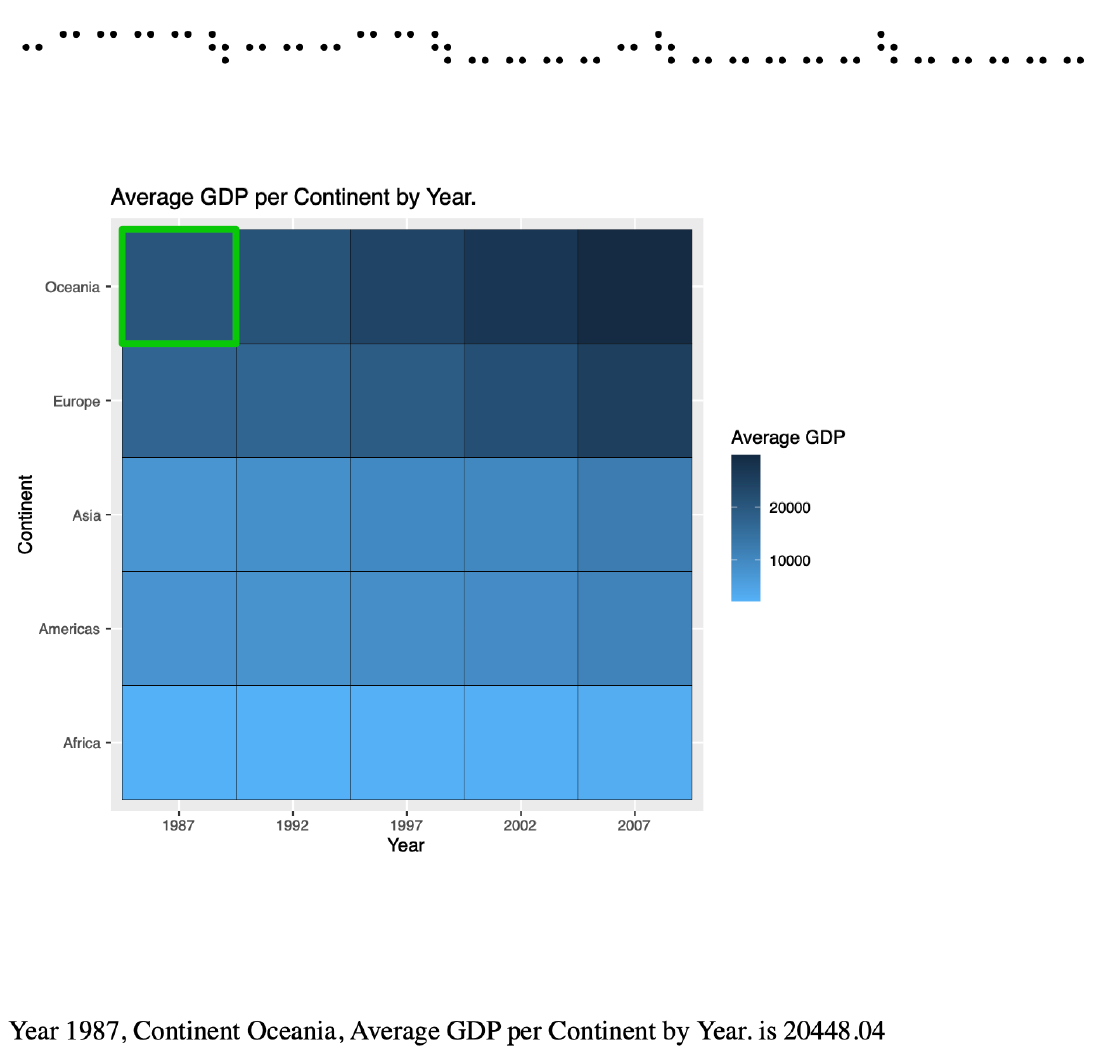}
  }\\[1ex] 
  \subcaptionbox{\textcolor{black}{Box plot}\label{fig:braille_verbose_c}}[.49\linewidth][c]{%
    \includegraphics[width=.85\linewidth]{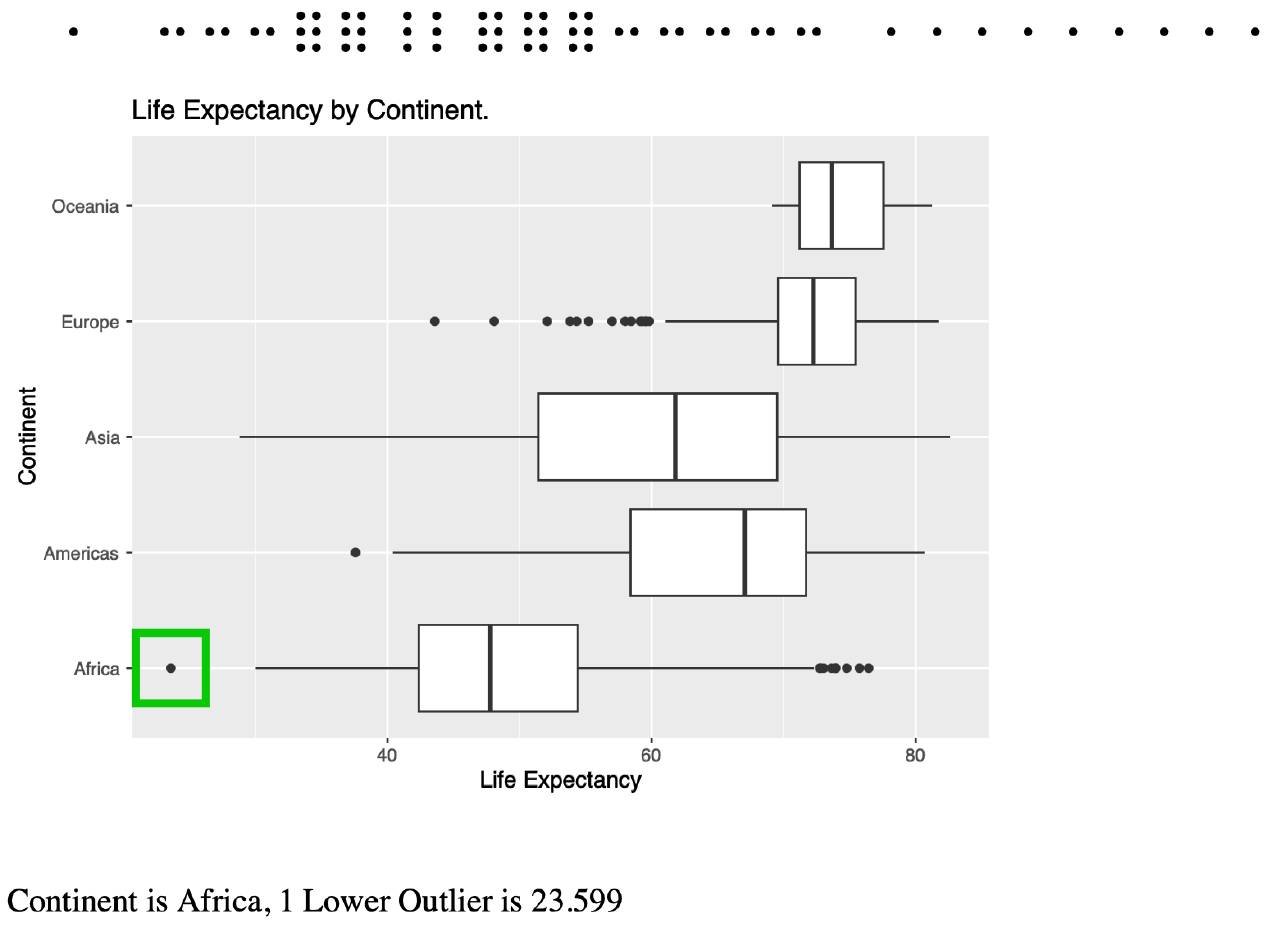}
  }\hfill
  \subcaptionbox{\textcolor{black}{Scatter plot}\label{fig:braille_verbose_d}}[.49\linewidth][c]{%
    \includegraphics[width=.9\linewidth]{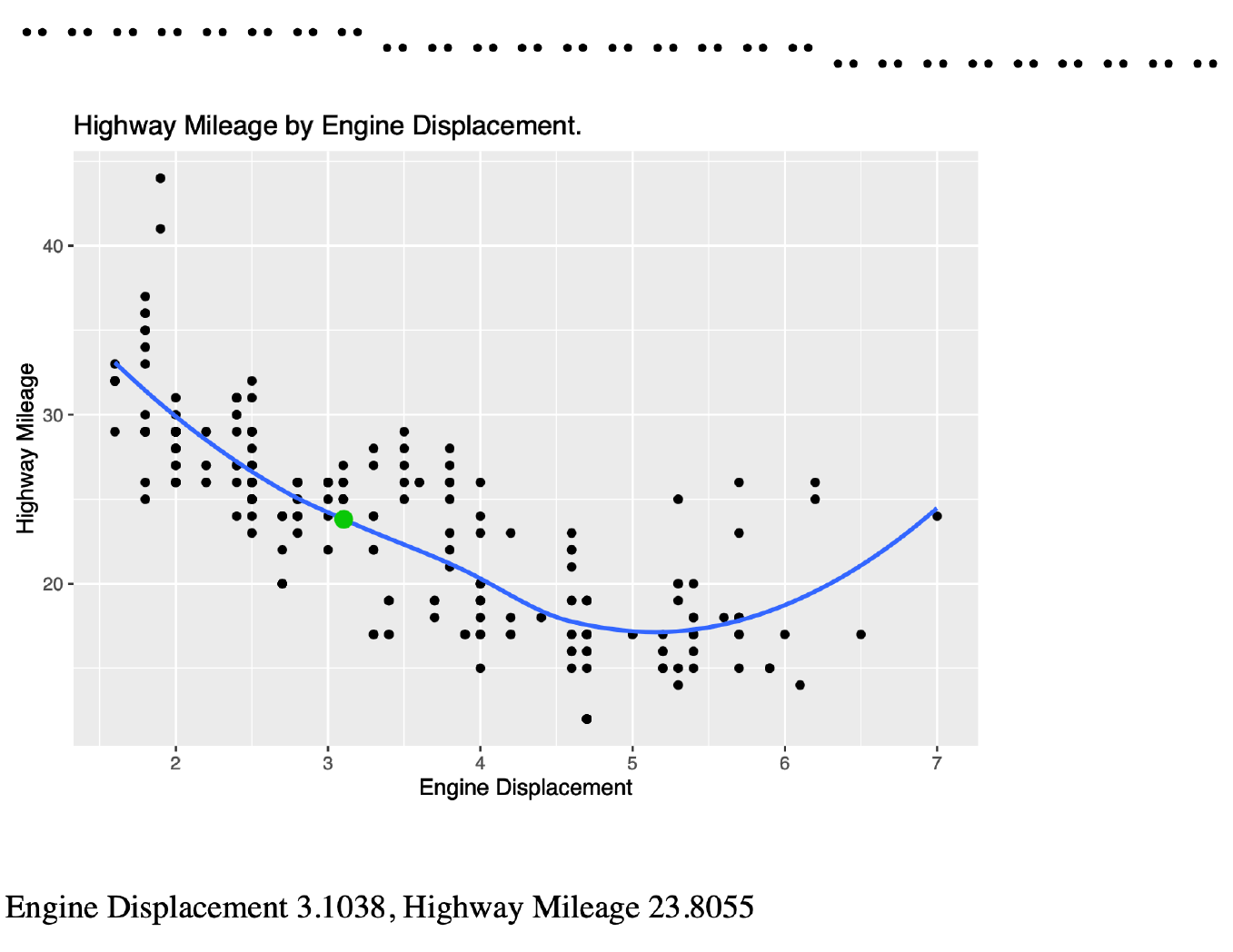}
  }
  \caption{Visualizations and their corresponding braille patterns with both braille and text modality (verbose mode) activated.}
  \label{fig:braille_verbose}
  \Description{This figure consists of four subplots. In the top-left is a bar plot representing the number of diamonds by cut. The braille pattern conveys the bar's value, complemented by a text description reading, 'Cut is Fair, Count is 1610.' In the top-right corner, there is a heatmap showing the average GDP per continent by year. The current focus is on the first element, with accompanying text stating, 'Year 1987, Continent Oceania, Average GDP per Continent by Year is 20448.04.' On the bottom-left is a series of box plots. The spotlight is on the box plot for Africa, with braille patterns above and text elaborating, 'Continent is Africa, 1 Lower Outlier is 23.599.' Finally, in the bottom-right corner, there is a scatterplot with the current emphasis on the line layer. Braille patterns depict the best fit line, and the text specifies, 'Engine Displacement 3.1038, Highway Mileage 23.8055.}
\end{figure*}

\bpstart{Bar plot} For braille-based bar plot representations (Figure~\ref{fig:braille_verbose_a}), each data value gets translated into specific braille characters based on its proportional magnitude. We use 8-dot braille's four distinct height levels to encode value ranges: braille chord 78 \includegraphics{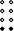} for 0-25\%, braille chord 36 \includegraphics{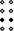} for 26-50\%, braille chord 25 \includegraphics{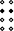} for 51-75\%, and braille chord 14 \includegraphics{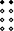} for 76-100\%.

\bpstart{Heat map} To make two-dimensional heat map data compatible with a one-dimensional single-line RBDs (Figure~\ref{fig:braille_verbose_b}), we use a row separator, encoded as braille chord 1256 \includegraphics[alt={braille pattern dots-1256}]{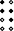}. This separator aids users in distinguishing between rows. For encoding cell values, we utilize only the first six dots of the 8-dot braille cell. Intensity levels for each cell are encoded as follows: braille chord 36 \includegraphics[alt={braille pattern dots-36}]{braille/braille_36.pdf} for 0-33\%, braille chord 25 \includegraphics[alt={braille pattern dots-25}]{braille/braille_25.pdf} for 34-66\%, and braille chord 14 \includegraphics[alt={braille pattern dots-14}]{braille/braille_14.pdf} for 67-100\%. Empty braille cell \includegraphics[alt={braille pattern null}]{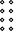} signify null or missing data.

\bpstart{Box plot} The braille representation of a boxplot (Figure~\ref{fig:braille_verbose_c}) uses specific braille characters that visually resonate with the corresponding parts of the boxplot. The encoding follows: braille chord 2 \includegraphics[alt={braille pattern dot-2}]{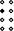} for both the lower and upper outliers, braille chord 25 \includegraphics[alt={braille pattern dots-25}]{braille/braille_25.pdf} for the whiskers,braille chord 123456 \includegraphics[alt={braille pattern dots-123456}]{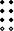} for the interquartile range box area (Q1-Q2 and Q2-Q3) respectively, and braille chord 456 \& braille chord 123 \includegraphics[alt={braille pattern dots-456}]{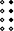}  \includegraphics[alt={braille pattern dots-123}]{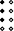} for the median. The extent of each section in the boxplot, represented by the count of braille characters, gives users an idea of the distribution of data within that section.

\bpstart{Scatter plot} Due to the constraints of single-line braille displays, our braille representation for scatter plots (Figure~\ref{fig:braille_verbose_d}) concentrates on the predictive or ``best-fit" line, forming the second layer of the scatter plot. \rv{For the test materials of our user studies, we used locally estimated scatterplot smoothing (loess) line as the predictive line because we thought it may convey more nuanced up and down trend patterns and point distribution than the straight linear regression line. However, this is not a limitation of the MAIDR system, as creators can also define their own predictive line data points using the MAIDR JSON schema.} The algorithm for this representation is similar to that of the bar plot but uses the predicted line's Y-values as the range for encoding. The points that traditionally form the primary layer of a scatter plot are difficult to represent within the confines of a single-line RBDs. However, the trend of the predictive line can be conveyed in some ways using braille. The encoding scheme is as follows: braille chord 78 
 \includegraphics[alt={braille pattern dots-78}]{braille/braille_78.pdf} for 0-25\%, braille chord 36 \includegraphics[alt={braille pattern dots-36}]{braille/braille_36.pdf} for 26-50\%, braille chord 25 \includegraphics[alt={braille pattern dots-25}]{braille/braille_25.pdf} for 51-75\%, and braille chord 14 \includegraphics[alt={braille pattern dots-14}]{braille/braille_14.pdf} for 76-100\%.


\subsubsection{Text}

In contrast to earlier approaches that offer pre-interpreted natural language descriptions or summaries of visualizations, the Text modality in our MAIDR system takes a different tack. It serves as a verbal commentary provided by a screen reader's speech synthesizer, focusing on individual data points currently under the user's scrutiny. This design choice aligns with our objective to enable blind users to actively explore and derive their own insights from the data, rather than passively consuming interpretations made by others. This approach is supported by Lundgard and Satyanarayan's research \cite{lundgardAccessibleVisualizationNatural2022}, which emphasizes the importance of human-driven exploration for higher-level perceptual trends and contextual insights in data visualization interpretation.

To cater to varying user needs \rv{(\textbf{DC5})}, we provide two modes within this Text modality: Terse and Verbose, as detailed in Table~\ref{table:text_verbose_terse}. In Verbose mode, the system generates comprehensive descriptions, which include x and y labels and their corresponding values, offering a nuanced understanding of the data point in focus. Alternatively, the Terse mode provides succinct, essential information, generally limited to the x and y values, for quicker navigation and interpretation.

\begin{table*}[ht]
  \caption{Text Verbose \& Terse in the MAIDR System Across Various Visualizations.}
  \label{table:text_verbose_terse}
  \footnotesize
  \begin{tabular}{p{1.5cm} p{4cm} p{4cm} p{4cm}}
    \toprule
                                & {Text:Verbose}                                                              & {Text:Terse (Horizontal Navigation)} & {Text:Terse (Vertical Navigation)} \\
    \midrule
    {Bar plot}                   & Cut is Fair, Count is 1610                                                  & Fair, 1610                           & -                                  \\
    \midrule
    {Heat map}                   & Year 1987, Continent Oceania, Average GDP per Continent by Year is 20448.04 & 1987, 20448.04                       & Oceania, 20448.04                  \\
    \midrule
    {Box plot}                   & Continent is Africa, 1 Lower Outlier is 23.599                              & 1 Lower Outlier is 23.599            & Africa, 1 Lower Outlier is 23.599  \\
    \midrule
    {Scatter plot (Point Layer)} & Engine Displacement 1.6, Highway Mileage [29,32,33]                         & 1.6, [29,32,33]                      & -                                  \\
    \midrule
    {Scatter plot (Line Layer)}  & Engine Displacement 1.9051, Highway Mileage 31.3885                         & 31.3885                              & -                                  \\
    \bottomrule
  \end{tabular}
\end{table*}

\begin{table*}[ht]
  \caption{Overview of Participant Demographics, Refreshable Braille Displays, Gender, Ages, Self-Reported Knowledge of Statistical Visualizations, and Screen Readers Employed in the Study.}

  \label{tab:participant}
  \scriptsize
  \begin{tabular}{p{0.4cm} p{2.5cm} p{1.1cm} p{0.4cm} p{1.2cm} p{3.2cm} p{0.8cm} p{1cm} p{0.8cm} p{1cm} p{1.4cm}}
    \toprule
    PID & Braille Display Name       & Gender     & Age & Education & Major                              & \multicolumn{4}{c}{Self-Reported Visualization Knowledge} & Screen Reader                                \\
    \cmidrule(r){7-10}
        &                           &            &     &           &                                    & Bar plot                                                   & Heat map       & Box plot & Scatter plot &      \\
    \midrule
    P01 & Mantis Q40                & Non-Binary & 27  & Master    & Journalism                         & 1                                                         & 1             & 1       & 1           & NVDA \\
    P02 & Active Braille 40         & Female     & 45  & Master    & Education                          & 1                                                         & 0             & 1       & 1           & JAWS \\
    P03 & Brailliant BI 40X         & Male       & 35  & PhD       & Applied Statistics                 & 1                                                         & 1             & 1       & 1           & JAWS \\
    P04 & Orbit reader 40           & Female     & 18  & High School & Not Applicable                   & 1                                                         & 0             & 1       & 1           & JAWS \\
    P05 & PAC Mate 20               & Female     & 57  & Bachelor  & Sociology and English              & 1                                                         & 1             & 0       & 1           & JAWS \\
    P06 & Mantis Q40                & Female     & 57  & PhD       & Computer Based Music Theory        & 1                                                         & 0             & 1       & 1           & NVDA \\
    P08 & BrailleNote Touch Plus 32 & Male       & 22  & High School & Not Applicable                   & 1                                                         & 0             & 1       & 1           & JAWS \\
    P09 & Braille EDGE 40           & Non-Binary & 25  & Bachelor  & Communication and Spanish          & 1                                                         & 0             & 1       & 1           & NVDA \\
    P10 & Focus 40                  & Female     & 48  & PhD       & Anthropology                       & 1                                                         & 0             & 1       & 1           & JAWS \\
    P11 & Brailliant BI 32          & Male       & 33  & Master    & Electrical and Computer Engineering & 1                                                         & 1             & 0       & 1           & Orca \\
    P12 & HIMS QBraille XL 40       & Female     & 26  & PhD       & Cognitive Neural Science           & 1                                                         & 0             & 1       & 1           & JAWS \\
    \bottomrule
  \end{tabular}
  \newline
  \raggedright{\textit{Notes: 1) The digits in the braille device name refers to the number of braille cells it has. 2) The ``1" and ``0" Self-Reported Knowledge represent whether or not the participant is familiar with the visualization.}}
\end{table*}

\subsubsection{Sonification}

The system integrates sonification as a form of non-verbal auditory feedback. Grounded in established sound mapping strategies that use stereo panning for x-axis values and pitch levels for y-axis values~\cite{summersAccessibleDataVisualizations2014,huntInteractiveSonification2011}, our system extends these methodologies with new implementations. This panning-pitch mapping approach is straightforward for two-variable visualizations like bar plots and the line layer in scatter plots.

For heat maps, we introduce a novel correlation between sound pitch and cell intensity (Figure~\ref{fig:braille_verbose_b}). Higher tones are associated with darker colors (higher values), and lower tones signify brighter colors (lower values). We also offer the option of four-channel surround sound for horizontal and vertical axes, although vertical navigation defaults to mono audio on standard two-channel devices.

In box plots, we innovate by adding different sound textures to represent various statistical elements. The interquartile range employs a double-tone, whiskers feature a single-tone, and outliers are denoted by incremental sounds. Empty outlier cells trigger a non-disruptive beep as an additional feature.

In scatter plots with multiple y-values for a single x-value, our system offers two distinct sonification strategies: (1) \textbf{Sonification Combined} plays the sounds representing all associated y-values simultaneously, and (2) \textbf{Sonification Separate} introduces time intervals between the sounds to facilitate clearer comprehension of the distribution of y-values for a given x-value.

\subsubsection{Review}
\label{sec:review_mode}

The Review feature serves as an auxiliary modality that provides both verbal and tactile representations of the currently focused data point, complementing the T mode. For screen reader users, although data values are vocally presented through a speech-to-text modality, the corresponding braille display often only shows this information transiently, similar to a flash message. This limitation arises because dynamic text updates are typically accomplished using `aria-live' alerts, and not all screen readers convert these transient messages into braille. Consequently, users who rely on braille are challenged to retain large or complex data values in their short-term memory, as these are not persistently displayed. This challenge was identified in our user studies, and we implemented the Review feature in direct response to user feedback \rv{(Section~\ref{sec:finding_braille_power})}.

When review mode is activated, it opens a pop-up overlay that covers the current navigation focus. Like the primary modalities, Review can also be toggled on or off. When activated, the overlay hides the Braille modality and exclusively displays the currently focused data value in literary braille on the user's RBDs. \rv{
We attempted to display both non-verbal B and verbal R modalities simultaneously on the braille display. However, due to the limited number of braille cells on a single-line RBDs, we were unable to fit both modalities in one line. This design decision was made to accommodate the limited cell availability in single-line RBDs, which have been conventionally adopted by the blind community.
}

%% file: 5-user-study.tex
\section{User Studies}
\label{sec:study}

The research plan received approval from the University's Institutional Review Board (IRB) before running the study, and we adhered to all necessary protocols during the research. 

\subsection{Participants and Apparatus}
\label{sec:participant_apparatus}

To recruit blind participants who might have a foundational understanding of statistics and the four visualizations covered by the MAIDR system, we began by sending out a sign-up form through two blind communities: the National Federation of the Blind (NFB) and Program-L (a discussion group for blind programmers). 



Of the 176 individuals who filled out the form, we narrowed down our pool by excluding those who did not self-identify as totally blind, did not have required software and hardware apparatus, such as single-line RBDs and screen reader, who were only familiar with two or fewer of the visualizations essential to our study, or lacked understanding of statistical concepts, such as correlation. As a result, we contacted 18 potential candidates. However, only 11 (Table~\ref{tab:participant}) managed to complete the interview, with others either not responding to our follow-up communications or withdrawing due to scheduling conflicts.

The average age of the 11 participants was 35.73 (\textit{SD} = 13.92), with ages spanning from 18 to 57. The demographic breakdown revealed 8 white individuals, 2 Asians, and 1 African American. In terms of gender distribution, the group included 6 females, 3 males, and 2 individuals who self-identified as non-binary. Their academic qualifications varied, with some having just completed high school, while others held bachelor's, master's, or PhD degrees. They also came from various academic backgrounds, such as Statistics, Journalism, Special  Education, Computer-Based Music Theory, Symbolic Systems, and Electrical Engineering.

Regarding their visual impairments, a majority of them had been blind since birth because of gene-related disease. Three participants developed visual impairments after the age of 10. Specifically, participant P10 became totally blind due to a brain tumor, while participants P05 and P08 faced visual loss as a result of retinal detachment. When inquiring about their familiarity with braille display, on average, participants began learning Braille at 7.77 (\textit{SD} = 4.45) years of age, with some starting as early as 4 and others as late as 16. The majority (\textit{n} = 7) were introduced to Braille in regular public schools, while a smaller group (\textit{n} = 4) received their instruction in specialized institutions. To optimize their experience with the MAIDR system, we gathered data on their preferred screen readers  and operating systems beforehand. Seven participants rely on Job Access With Speech (JAWS) for Windows, three use NonVisual Desktop Access (NVDA) on Windows, and one employs Orca on a Linux platform. It is worth mentioning that P07, who isn't among the 11 participants and only completed half of the study, utilizes VoiceOver on MacOS.


\renewcommand\arraystretch{1.3}

\subsection{Procedure}
\label{sec:procedure}

Before the user study, participants were asked to sign a consent form, schedule a meeting using Calendly, and ensure their refreshable braille display works. All sessions were conducted via Zoom meetings, typically lasting between 2.5 to 3 hours. Participants were requested to share their screens and grant us authorization to record their activities during the study session. Upon concluding the session, participants were ask to download the JSON log file from our MAIDR system that recorded their keyboard events and send it to us via email. As a small token of appreciation for their time and effort, they received a \$30 Amazon gift card.

Participants started the session by answering several questions related to their background, such as personal and demographic details, education, visual impairments, and familiarity with braille, as outlined in Section~\ref{sec:participant_apparatus}. Following this, participants were asked to share the screen. The researchers then assisted them in accessing the user study website where they can download screen reader dictionary files. \rv{We prepared dictionary files tailored to various screen readers like JAWS, NVDA, and VoiceOver. The purpose of this dictionary file was to prevent our users from being interrupted by the screen reader speech when moving their cursor in non-verbal braille (B) mode. Since most screen readers pronounced unicode braille characters by default, users would hear the screen reader speech for the braille chord when B mode is toggled on. This was not ideal in our user study settings because users had to hear other verbal (i.e., text mode) and non-verbal (i.e., sonification) sound, depending on their  modality combination. With the speech dictionary files, we were able to mute the screen reader speech for the unicode braille chord. We provided instructions on how to enable eight-dot braille for their screen reader and RBDs as the MAIDR system required eight-dot braille unicode to maximize the resolution. For JAWS users, we asked them to disable flash message braille for aria-live speech to maintain the same condition as NVDA users. This was necessary because JAWS users would receive the aria-live speech message from text mode in verbal braille on their braille displays as a temporary flash message when moving their cursor around each data point, which was not the case for NVDA users.\footnote{After our user studies, NVDA was updated to support dynamic aria-live message in flash braille (2023.2 or later).} Further details on user study pre-setup are provided in Appendix~\ref{sec:appendix_presetup}.}

Participants completed four sections, each dedicated to a distinct visualization. Each section included a tutorial, task, survey, and follow-up interview.

\bpstart{Tutorial} For every visualization, participants were asked about their prior knowledge of the visualization, such as whether they've encountered it before or if they've had any tactile experiences with it. Following that, participants were trained on how to use the MAIDR system for that specific visualization. This encompasses keyboard navigation, toggling various modalities on and off, and auto-playing sonification based on their desired direction, among other features (instruction can be found in Appendix~\ref{sec:appendix_tutorial}).

\bpstart{Task} Once we've confirmed with participants that they are well acquainted with the MAIDR system through the tutorials, they were directed to the task session. Each task session included 4-6 questions. To avoid overlapping the interviewer's voice with the screen reader's sound, participants were asked to start the question only after receiving a ``start" cue from the researchers. Upon the completion of a question, participants said ``done," at which point the researchers can interject or move to the next question.

\bpstart{Survey} After completing the task session, participants were asked to take a survey. We guided participants through the System Usability Scale (SUS) questions \cite{SUSQuickDirty1996,brookeSUSRetrospective2013}, and asked them to rate the MAIDR system's usability for each visualization on a scale of 1 (strong disagreement) to 5 (strong agreement).

\bpstart{Interview} Following the survey, participants were asked several open-ended questions. These questions delved into topics such as the usefulness of the modalities supported by the MAIDR system, their overall approach to utilizing the various modalities for the given visualization, which modalities they'd prefer to have enabled by default, any limitations they identified, and any additional features they'd like to see incorporated.

\subsection{Dataset and Tasks}
Gapminder dataset \cite{gapminder} was used for the tasks. The rationale behind this decision was twofold. First, Gapminder, as an open-source dataset, features globally relevant variables such as continent, life expectancy, year, total population, and GDP. These variables can be readily interpreted by the general public, without requiring specialized background knowledge. Second, Gapminder seamlessly offers the variety of data variables necessary for our study. It covers both numerical variables, such as GDP and Life Expectancy, as well as categorical (i.e., factor) variables like Continent and Year. This rich variety of variables guaranteed that our four visualizations were created and tested within MAIDR system.

Our task questions (see Appendix~\ref{sec:appendix_taskql}) for each type of visualization were designed to see how well users can get insights from the data using the modalities provided by the MAIDR system. \rv{Drawing on the eleven characterization labels and the four prevalent categories of data visualization inquiries (namely, lookup, compositional, visual, and non-visual) suggested by Kim et al.  \cite{kimAnsweringQuestionsCharts2020, kim2023exploring}, we formulate a series of questions pertinent to each visualization. Subsequently, we refine this set to 4-6 questions, taking into account the time constraints and the distinct statistical concepts emphasized by the visualizations, as outlined in \cite{oyana2020spatial}.
Therefore, for the bar chart, we formulated two ``find extremum" lookup questions, two compositional questions on value comparison, and one visual question centered on the chart's shape. In the case of the heatmap, we included two ``retrieve values" lookup questions, one ``correlate" lookup question, and three visual questions concerning the heatmap's spatial information and patterns. Since box plots are essential for illustrating statistical concepts like quartiles, whiskers, and outliers, we incorporated two compositional questions for statistical value comparison, one ``find anomalies" question to identify the continent with the most lower outliers, and one visual question to deduce the skewness of the boxplot. For the scatter plot, we focused on the correlate aspect between variables, highlighting correlation and overall ``data distribution", which led us to include two ``correlate" compositional questions and two visual questions about line trends and point distributions.
}

\subsection{Data Analysis}


In our study, we gathered data through audio and screen recordings during task sessions and subsequent interviews. Additionally, we collected low-level action logs from participants. 
By comparing these logs with the recordings from the study session, we were able to determine the sequence in which participants used different modalities during their interaction with each visualization for various questions. For example, the sequence ``Start $\rightarrow$ TB $\rightarrow$  TBR $\rightarrow$ TB" for the first question related to the box plot indicates that the participant initially activated the `TB' modalities before navigating, then added the `R' modality, and later turned off 'R.' Using such individual interaction patterns as a foundation, we created aggregated state diagrams \cite{kieras1983generalized, wasserman1985extending}. 



To delve deeper into the insights conveyed by the state diagrams, we analyzed the interview data transcribed from audio. We employed an inductive coding approach \cite{thomas2006general} to categorize the data. Two researchers independently embarked on an open coding exercise, with interviews from three participants. Their objective was to formulate a consistent coding framework or ``code book." One researcher continued with the coding process for the remaining transcripts, maintaining a regular dialogue with the second to ensure coding accuracy and consistency. In instances where coding interpretations diverged, the input of a third researcher was sought. Upon finalizing the coding process, the team collaboratively synthesized the individual codes, grouping them into overarching themes pertinent to addressing RQ1 and RQ2.

%% file: 6-findings-RQ1.tex
\section{Findings}

We evaluated the usability of our MAIDR system with 11 blind participants using the System Usability Scale (SUS) test \cite{SUSQuickDirty1996,brookeSUSRetrospective2013}. On average, bar plots scored 81.36 (\textit{SD} = 17.00), heat maps scored 75.5 (\textit{SD} = 18.36), box plots scored 74 (\textit{SD} = 17.37), and scatter plots scored 70.25 (\textit{SD} = 21.49). These usability scores appear to correlate with the inherent complexity of each type of visualization, suggesting that the results may also reflect the learnability of each visualization. While the SUS test offers a quick snapshot of our system's usability, it does not capture all nuanced aspects. Below, we delve deeper into the user experience.

\subsection{Different Strategies of Utilizing Multiple Modalities to Explore Statistical Visualizations}
\label{sec:strategies}

To better understand the varied strategies employed by blind users when exploring statistical visualizations, we examined the system logs detailing participants' usage of multiple modalities to generate an aggregated state diagram for each visualization. By cross-referencing this with the quotes from participants' interviews, we gained insights into their interaction behaviors, from general strategies across four visualizations to preferred strategies for exploring specific visualizations.

\subsubsection{Diverging Approaches to Modality Engagement: Whole-to-Part vs. Part-to-Whole Strategies} \hfill

\bpstart{Whole-to-Part Strategy: Initial Global Scanning Followed by Detailed Analysis}
A majority of our participants leaned toward employing a "whole-to-part" strategy. This entails using either sonification or braille initially to grasp a global view of a graph or dataset before delving into the finer details. P02 highlighted the value of this approach, saying, ``\textit{if I give you a comparison between [sonification and] braille which I typically teach. We look at an overall view of a graph. You at least have that part in place where you can listen to and visually see, or tactually see, the overall graph before you start looking at the details, which is a good thing.}" P03 specifically emphasized the utility of sonification for quickly scanning shapes in bar charts, noting, ``\textit{This sonification, I think, is good for a quick scan of the shape of the bar chart. For example, I think one of the questions you asked was, which continent has the highest population so, [it is] very easy to skim across and locate the highest bar with the sonification.}" P06 also mentioned that text becomes essential only when specific details are necessary, explaining, ``\textit{So it probably depends on the nature of that. I might enable sonification and braille, and that, and then do text as an option. Because text is the one that's kind of only useful when you're at the point where you need to be.}"

\bpstart{Part-to-Whole Strategy: Detail-First Approach Contrary to the Whole-to-Part Method}
On the contrary, P04 exhibited a preference for a "part-to-whole" strategy, starting with textual elements for initial familiarity with a graph before utilizing braille and sonification for more in-depth exploration.
P04 elaborated, ``\textit{I like the text for just getting familiar with the graph, like the y-axis and the x-axis you know. And the title, and whatnot. And once you get that, I think you can use the braille and the sonification to get the shape.}" They further noted, ``\textit{I'd say the text as a default. I mean, I think, just to get familiar with the graph, I think. Oh, doing like a quick overview of all the axes what they're asking for and then once you get familiar with that, then you can start enabling braille and the sonification to look even more in-depth.}"

\subsubsection{Modality Preferences: Influences of Individual Skills and Challenges} \hfill

\bpstart{Challenges of Sonification for Users with Hearing Impairments}
Our findings indicate that sonification, while generally useful, may not be the optimal choice for all users. P02, who has a high-pitch hearing impairment, explicitly stated that text and braille are more useful for them than sonification. They noted, ``\textit{I get a lot more from text than I do from the sonification part, just because of the, having the hearing impairment. Because my hearing impairment is higher pitch. So the higher the pitch, the less I hear.}" P02 further elaborated on their preference, stating, ``\textit{for me it's going to be between the braille and the text, since i'm not really hearing. I can hear the pitches, but they're not, they don't clue me in to what I'm looking at .... honestly I don't think there needs to be one out of it because it gives the user a choice. I mean, my preference is, you know, braille text, but somebody else may be sonification with the braille.}"

\bpstart{The Particular Value of Braille for Certain Users}
Several participants expressed a strong preference for braille over other modalities. P02 mentioned, ``\textit{I think Braille is probably, to me, most useful. But if you need data information or anything. Then the text pieces are gonna help.}" P08 added their perspective, stating, ``\textit{For me, I think, um, with topographics, I use topographics somewhat frequently, relatively frequently, I would say so. It's like, I think, just using my hands to feel a shape is something that I'm used to and something that is a really quick way for me to get an understanding of an overall shape, with a little bit more granularity than sonification because sometimes sonification is a little bit unclear.}"

\bpstart{Sonification's Value for Users with A Strong Sense of Pitch}
On the flip side, P06, who has a strong sense of relative pitch, found sonification particularly useful. They said, ``\textit{No [I don't have perfect pitch]; but I'm very good at relative pitch. So I work in computer music and I'm used to listening to very small frequency deviations.}"

%% file: 7-findings-RQ2.tex
\subsection{Best Practices for Integrating Braille into Accessible Statistical Visualizations}
\label{sec:practices}
The potential of \rv{non-verbal} braille in enhancing accessible statistical visualizations is a relatively uncharted domain. As highlighted by feedback from our study participants, there's a notable intrigue and commendation regarding this pursuit. P10 revealed the novelty in encountering such an integration, expressing, ``\textit{That's interesting, because I think more people are working on sonification than are working on braille. So this is the first time I've encountered someone doing a braille shape.}" Similarly, the efforts and challenges of matching statistical visualizations with single-line braille displays didn't go unnoticed. P12 candidly remarked, ``\textit{It's so tough to do what you guys are doing with single line braille displays, and not a lot of people try to tackle this.}" In light of these insights, this section endeavors to synthesize the best practices and lessons we derived from our research, aiming to provide a structured approach for future endeavors in braille-integrated statistical visualizations.

\subsubsection{The Unspoken Power of Braille in Non-Verbal and Verbal}\hfill
\label{sec:finding_braille_power}

\bpstart{Importance of Verbal-Tactile Representation in Refreshable Braille Displays} Our user studies highlighted the need for a reliable, consistent display of verbal data values on refreshable braille displays. Traditional screen readers often employ a `flash message' feature to momentarily present dynamic aria-live text, serving as a verbal data point, on braille displays. However, the availability and behavior of this feature vary across platforms, creating inconsistencies that can impede effective data comprehension. Notably, the NVDA screen reader lacked this `flash braille message' feature as of our study, a shortcoming explicitly called out by P01:\textit{``If there is a way to make it, that's I don't know through an add on, or whatever. That NVDA can have the choice to display the text. That will be very nice for a braille reader like me ...''}
Even when flash messages are available, they are ephemeral and may disappear before the user has had a chance to fully read and internalize the data, for example, as pointed out by P02:
``\textit{The \rv{[verbal]} braille \rv{[for data values]} kept disappearing, and that's what I was having a hard time, when I said I couldn't see it."} 


\rv{
In response to feedback from P1, P2, and P3 about the importance of reading data at their own Braille speeds, our team developed an auxiliary review mode that displays verbal Braille on users' RBDs (Section~\ref{sec:review_mode}), and this feature was then introduced to P4 through P11.
}
Our user-driven implementation of a dedicated ``Review Modality" successfully addressed these challenges \rv{highlighted by P1-3}. According to feedback, users found it particularly useful to switch between sonification and verbal-tactile review for a more holistic understanding of the data. P08 noted how the Review Modality was easier to comprehend than a stream of spoken digits (the verbal-auditory text mode):``
  \textit{I think the review mode was really useful, yeah, I think reading the different numbers, JAWS would read it as a stream of digits. So for me it was easier to comprehend using braille."} P04 found the verbal-tactile representation to be highly effective, especially when compared to fast text-to-speech output:
\textit{"I couldn't really read it [text], like I could read it with JAWS. It was a little too fast, especially when looking at numbers."}
Thus, the stable and non-disappearing tactile representation offered by the Review Mode not only improves accessibility but also enhances the user experience in comprehending complex data.

\bpstart{The Value of Non-Verbal Braille Modality} 
Non-verbal braille patterns, despite their inherent limitations, stand as a vital instrument in enhancing accessibility. Primarily, they serve as a tool for holistic comprehension, facilitating users in grasping overarching patterns and trends without getting bogged down in the specifics. P02 encapsulated this advantage --- ``\textit{I appreciate what you've done so far with this. It is knowing that you can take something that you know has more than just one line of text, and make it fit to a braille display where people can actually see the overall trend of something versus going into more detail.}" This ability to perceive the larger picture can be particularly beneficial in contexts where a broad understanding is more crucial than dissecting individual data points.

\subsubsection{Limitations and Potentials of Braille in Different Types of Data Visualization}
While the introduction of non-verbal tactile representation is groundbreaking, the current constraints of single-line braille displays limit its full potential. One evident constraint is its limited granularity. The confinement of the braille cell hinders its ability to represent diverse levels, as P01 mentioned, ``\textit{I think it's a little limited by just the fact that, like the cell only has 3 levels that you can put in the graph. To be honest, it's kind of like nice, but it's not going to give you the degree of specificity that you would need from other things, I think.}" P12 and P08 further highlighted the challenge, noting the similarities in braille dots, emphasizing the limitation in variety.

Besides, several participants found the cognitive load considerably high when interfacing with a single-line display. This sentiment is effectively captured by P05 --- ``\textit{It is a little bit complicated because we only have one-line braille displays right now. If we had more multiple line braille displays, that would be much better.}" This feedback underscores the imperative need for exploring multi-line braille displays to potentially alleviate cognitive burdens.

Moreover, the utilization of braille in conjunction with sonification and text for interpreting various types of graphs and data visualizations elicits a range of opinions among users. Braille's efficacy largely depends on the type of data representation at hand—be it bar plots, heat maps, box plots or scatter plots. \\

\bpstart{Braille in Bar Plots: Least Useful But Not Insignificant}
Participants like P11 and P12 highlight that braille tends to be the least useful when interpreting bar plots. P12 notes that because the heights of bars like ``braille chord 78 \includegraphics[alt={braille pattern dots-78}]{braille/braille_78.pdf}, braille chord 36 \includegraphics[alt={braille pattern dots-36}]{braille/braille_36.pdf}, braille chord 25 \includegraphics[alt={braille pattern dots-25}]{braille/braille_25.pdf}, and braille chord 14 \includegraphics[alt={braille pattern dots-14}]{braille/braille_14.pdf} in reality do feel somewhat similar,'' braille offers less differentiation. However, it doesn't mean braille is completely ineffectual as P11 indicated ``\textit{the least useful in this case, I would say it would be the one which I would have been able to do the question without, and that would have been braille. But it doesn't mean it (non-verbal braille) wasn't useful, it gives an overview."}
While it doesn't mean braille is completely ineffectual, P11 notes it ``gives an overview''. P12 notes that because the heights of bars like ``78, 36, 25, and 14 in reality do feel somewhat similar,'' braille offers less differentiation. \\

\bpstart{Braille in Heat Maps: A More Favorable Scenario}
When it comes to heat maps, braille's utility improves significantly. Multiple participants, including P11, find braille to be more helpful in heat maps than in bar plots, as they mentioned ``\textit{In this case [heat map] I think  the one I would reach for first here is braille. Once I understood how it works. because I'm able to see pretty much the whole heat map."} In addition, P01 mentions that braille allows them to quickly identify the ``highest value'' and ``eliminate the bottom 3 rows,'' emphasizing its role in efficient data skimming. However, the constraint of single-line RBDs was brought up by P04, ``\textit{I've used tactile graphics. It doesn't look like this and everything is different because this is a single line."}, suggesting that multi-line displays might enhance this utility. \\

\bpstart{Braille Strength in Box Plots: A Clear Winner}
Box plots bring out the strengths of braille representation, especially in interpreting statistical concepts like medians, skewness, and outliers. Participants like P04 and P05 specifically praised braille for its ability to convey the ``shape of the box and whiskers'' and the ``spatial layout of the actual overall plot.'' P10 and P11 pointed out that braille was not just about the data points but the distribution of those data points, adding another layer of understanding. The reason, as outlined by P10, is the alignment of Braille's characteristics with the fundamental components of a boxplot: ``\textit{I'd prioritize braille over text and sonification when dealing with boxplots. The braille representation felt more intuitive. Distinct braille markers, such as those for medians, offered more clarity than their sonification counterparts, which rely on pitch variations tied to data points.}" \\

\bpstart{Scatter Plots: A Challenge for Braille}
In scatter plots, braille struggles, mostly due to its limitations in representing multi dimensional data on a single line. While P04 thought braille could provide an ``initial shape,'', P03 ultimately found sonification to be more useful for understanding scatter plots ``\textit{sonification is very useful. Braille is probably, of the 4 examples you've got, that braille is probably the least useful here, but I would expect that. It is the limitation of a single line braille display.}
. P08 attributed this to braille's inability to effectively communicate ``inclines'' or ``declines.'' \\

\subsubsection{The Importance of Training and Scaffolding}

As P02 pointed out with respect to heat maps, ``\textit{I was trying to see what the braille was doing versus what I was hearing, and it wasn't matching. That's why I'm like, okay, that is not making sense to me. So. But now that I was just given the explained version of what the braille does. I'm like, okay, that makes sense.}'' This experience reflects that, akin to traditional visualizations, alternative tactile modalities like braille also necessitate a level of training and scaffolding to be effective.

P12 illuminated how prior experiences with different representations could shape current interpretations, particularly in a single-dimensional setting. They said, ``\textit{Maybe that was just me (find this scatter plot hard to interpret), because, like I said, I'm used to seeing them in a like two-dimension, you know, format like on graph paper or something.}'' Moreover, P12 shared their thoughts on encountering a new tactile representation of a box plot: ``\textit{For box plots, I don't know that I've ever seen them represented tactilely so to learn a 1D representation wasn't counterintuitive, because I didn't know another way.}'' This accentuates the need for mindful design choices when introducing novel tactile data representations.

P03 raised a crucial point about the diverse learning curves associated with new tactile modalities. They noted, ``\textit{I'm in a fortunate position here, where I'm kind of statistically minded. And that for other people just the actual concepts, might be more, more challenging.}'' This highlights the essential need for comprehensive training programs and adaptive interfaces that cater to a range of users' background knowledge and experience.
The training programs should focus not only on familiarizing users with tactile modalities but also on building a foundational understanding of the statistical visualizations being depicted. As P05 remarked about encountering a box plot for the first time: ``\textit{I think it will require some learning and you cannot [get familiar with the box plot] with the tutorial in 5 mins. I understood when you described it. I understood what you were talking about, and how the outliers and things like that. That's not a problem. It's just that I have never seen a visualization of a box plot, I've never done.}'' This calls attention to the significant role of a comprehensive and multifaceted training approach that aids users in effectively navigating and understanding braille displays. This highlights that blind users need structured guidance to bridge the gap between traditional visual understanding and tactile interpretation.

%% file: 8-discussion.tex
\section{Discussion and Future Opportunities}
\label{sec:discussion}

\subsection{MAIDR: A System Manifesting Autonomy in Utilization of Multiple Modalities}
\label{subsec:diversity_strategies}

Previous research has underscored the pivotal role of autonomy in enabling blind users to fully comprehend and interpret data visualizations \cite{thompsonChartReaderAccessible2023}. Through our comprehensive analysis detailed in Section~\ref{sec:strategies}, it became evident that user preferences span a broad spectrum when it comes to activating distinct modalities in accessible visualizations. This diverse range highlights the individualized inclinations and levels of comfort users have for different visualization types. As they delve into a visualization with the aim of achieving a thorough understanding, distinct modality choices come to the fore. Notably, sonification and braille stand out, with each resonating with users for specific visualization types. For example, the tactile nature of braille seamlessly aligns with the structure of heatmaps and box plots. On the other hand, the auditory cues of sonification are more appealing when deciphering bar plots and scatter plots. By offering a range of modalities, our system endows users with a degree of autonomy, empowering them to align with a modality that best mirrors their individual needs and preferences when exploring various visualizations.

Delving deeper into the nuances of each modality, the importance of autonomy becomes even more pronounced, particularly when merged with navigation tools and advancements like RBDs. Catering to the varied requirements of our users, our system has been designed to offer adjustable braille display cell configurations (Section~\ref{sec:system}). The insights gathered from Section~\ref{sec:practices} reinforced our design choices, emphasizing the pivotal nature of such adaptability. Users frequently expressed their gratitude for the level of control our system offered, letting them tweak the cell numbers in tune with their needs, facilitating a bespoke experience during their engagement with accessible visualizations. On another note, while there were suggestions for additional features, certain users felt that just a single line of braille was adept at encapsulating the complete essence of a heatmap. Conversely, a few found adapting to the braille system slightly challenging, more so if they had prior exposure to real-time feedback from braille during navigation. Notably, feedback from participants indicated that enhanced autonomy could be advantageous for users with varying braille display cell counts, as this would allow them to personalize the cell number while navigating accessible visualizations.



\subsection{Affordances and Limitations of Braille Across Visualization Types}
\label{subsec:affordances_limitations}

\rv{Non-verbal} Braille, as a modality for representing visual information, has exhibited diverse efficacy when employed across varied types of visualizations. During the user study, we discerned that the utility of braille as a communicative tool differs considerably based on the nature of the visual representation. For instance, box plots proved to be an area where braille shines; its linear structure and comparative simplicity make it conducive to a braille interpretation. Users were able to grasp the data distributions and outliers effectively through the tactile medium.
Conversely, heat maps and scatter plots posed more challenges. Heat maps, with their two-dimensional data distributions, and scatter plots, with their point-based representations, are intrinsically more complex and dense. Translating this intricacy to the linear structure of braille proved to be less straightforward. The core challenge emanates from the innate constraints of the currently prevalent braille displays, which predominantly operate on a single-line format. This restriction can often truncate or oversimplify the rich information that visualizations (e.g., heat maps) aim to convey. For example, displaying heat maps with dimensions larger than 5x5 becomes highly challenging when limited to a single-line tactile format.

Such observations lead us to hypothesize that the evolution of braille display technology might hold the key to unlocking more comprehensive and nuanced interactions for the blind users. Multi-line RBDs, which could present information in a more expansive, grid-like format, seem to be a promising avenue. Such displays might enable a more faithful representation of visualizations by allowing for simultaneous tactile exploration across multiple rows or columns, thus potentially bridging the current gaps in accessibility and understanding.

\subsection{Future Opportunities}
\label{subsec:future_research}

We acknowledge several limitations inherent to the current iteration of our work. One of the most prominent challenges we faced revolves around the constraints of using a single-line RBDs. Given its limited cell capacity, it proved challenging to present both non-verbal (B) and verbal braille (R) modalities concurrently on the MAIDR system. This limitation impinges on the user's experience, compelling them to alternate between modalities rather than having an integrated experience. With future developments, we are keen on exploring potential solutions to this problem. One potential avenue is to divide the braille display, allocating dedicated portions to each modality. Such a configuration could potentially harmonize the presentation of both modalities, enhancing the user's interaction with the system.

\rv{
Future research can also invesigate how our MAIDR system can offer more nuanced user experience with when multi-line RBDs are used. We just tested our MAIDR system with a single-line RBDs in this study, recognizing its prevalence among blind users. However, as described in our design consideration (\textbf{DC2}), we designed our system to be adaptable to different braille display sizes. Users can select the appropriate column size for their device in the MAIDR system help menu (\textbf{DC5}; Section~\ref{sec:system_overview}). If a braille pattern exceeds the chosen size, it automatically wraps to the next line, allowing users to navigate between patterns using their braille display's scroll key. This approach guarantees that our system is seamlessly adaptable for use with braille displays of any size, encompassing various numbers of cells and lines, providing users who have access to more braille cells and lines the flexibility to adjust the granularity of the braille patterns they interact with. Our future work will explore how our MAIDR system can be adapted to multi-line RBDs and how it addresses the granularity issues that arise when using a single-line RBDs.
}

Another area for enhancement is the provision of more adaptable data analysis tools. As the data science field evolves, there is an increasing demand for accessible tools that offer blind users the liberty to undertake actions like data filtering and wrangling with ease. Such capabilities can greatly empower users, enabling them to tailor their data visualizations to their unique needs and preferences.
P03 emphasized the need for more flexible analytical tools, such as customizable scatter plot functionalities:
\begin{quote}
    \textit{``It would be nice if you could pick your own (smooth line) function rather than be constrained to something.''}
\end{quote}

Similarly, P05 points to the current system limitations in handling complex, multi-variable analyses, especially for topics that require comprehensive study: 

\begin{quote}
    \textit{``When you are doing data analysis. This (MAIDR system) will not do the full job...You will have to figure out how to do multiple things and that would be too much.''}
\end{quote}

Future work will address these user concerns by integrating MAIDR into more widely adopted data science and programming environments, such as R and Python. By doing so, we aim to extend the system's capabilities, making it a more versatile tool for advanced data analysis.


%% file: 9-conclusion.tex
\section{Conclusion}
\label{sec:conclusion}

In this paper, we introduced MAIDR, a multimodal system designed to make statistical visualizations accessible for blind individuals. 
\rv{Through interdependent co-design process within our mixed-ability team, we have proposed a system that extends beyond the traditional modalities of sonification and text descriptions by integrating braille and review modes. From} a user study involving 11 blind participants, we found that MAIDR not only facilitated accurate interpretation of various statistical plots but also allowed users to employ a range of strategies for combining multiple modalities.
Our results reaffirm the potential of using refreshable braille displays in the realm of accessible data visualizations, breaking away from the constraints of existing web technologies predominantly based on text and sound. The impact of MAIDR is best captured by the words of P03, a blind statistician: \textit{``Many thanks once again for showing me your system, it was a really great experience and I can't wait to use it in my everyday work. I really think it's a game changer for blind people in statistics and the sciences more widely.''}
The study also provides valuable insights into how user autonomy can be prioritized in design to allow for a personalized, engaging experience.

To further sustainable development and encourage dialogue, we have made the MAIDR system open-source, inviting collaboration at \url{https://github.com/xability/maidr}. We hope to inspire continued interdisciplinary efforts among HCI, accessibility, and data visualization communities to bring about meaningful change in how we approach the design of accessible visual technologies.

%% file: 10-acknowledgement.tex
\begin{acks}
This study was made possible by generous funding support from the Wallace Foundation Grant, awarded in conjunction with the first author’s 2022 ISLS Emerging Scholar award, the Institute of Museum and Library Services (IMLS) through the Laura Bush 21st Century Librarian Program (grant \#RE-254891-OLS-23), a 2023 TeachAccess grant, and a faculty startup grant. We also extend our heartfelt thanks to the National Federation of the Blind and Program-L communities for their invaluable participation and contributions.
\end{acks}

%% file: 11-appendix.tex
\setcounter{equation}{0}
\setcounter{section}{0}
\setcounter{figure}{0}
\setcounter{table}{0}
\setcounter{page}{1}
\makeatletter
\renewcommand{\theequation}{S\arabic{equation}}
\renewcommand{\thefigure}{S\arabic{figure}}

\newpage
\section*{Appendix}

\section{JSON Schema}
\label{sec:appendix_json}

\begin{lstlisting}[caption={JSON Schema for box plot in MAIDR. \textbf{Notes:} The \textbf{type} denotes the type of the plot with supported values such as `bar', `heat', `box', `scatter', and `line'. The \textbf{id} is the identifier added as an attribute of the main SVG. An optional \textbf{title} can specify the plot's title. The \textbf{axes} contain information about the plot's axes, with `maidr.axes.x.label' and `maidr.axes.y.label' providing axes labels. Either `maidr.axes.x.level' or `maidr.axes.y.level' will provide tick mark labels. Finally, \textbf{data} represents the main data for your plot.}, label=json-example, captionpos=b]
{
  type: 'box',
  id: 'myboxplot',
  title:'Highway Mileage by Car Class.',
  axes: {
    y: {
      label: 'Car Class',
      level: [
        '2seater',
        'compact',
        'midsize',
        'minivan',
        'pickup',
        'subcompact',
        'suv',
      ],
    },
    x: {label: 'Highway Milage'},
  },
  selector: document.querySelector(
    '#boxplot1 g[id^="panel"] 
    > g[id^="geom_boxplot.gTree"]'
  ),
  data: [
    {
      lower_outlier: null,
      min: 23,
      q1: 24,
      q2: 25,
      q3: 26,
      max: 26,
      upper_outlier: null,
    },
    {
      // etc
    },
  ],
}
\end{lstlisting}
\newpage
\section{Shortcut Keys}
\label{sec:appendix_shortcuts}

\begin{table}[ht]
  \caption{Keyboard shortcuts in MAIDR. The Control key is for Windows and the Command key is for Mac.}
  \label{table:navigation_shortcuts}
  \footnotesize
  \begin{tabular}{p{3.5cm} p{4cm}}
    \toprule
    \textbf{Function}                                         & \textbf{Key}                                  \\
    \toprule
    \texttt{Move around plot}                        & Arrow keys                              \\
    \midrule
    \texttt{Go to the very left, right, up, or down} & Control/Command + Arrow key                  \\
    \midrule
    \texttt{Repeat current focal point}              & Space                                     \\
    \midrule
    \texttt{Auto-play in the direction of arrow}     & Control/Command + Shift + Arrow key  \\
    \midrule
    \texttt{Stop Auto-play}                          & Control/Command                     \\
    \midrule
    \texttt{Auto-play speed up}                      & Period  \\
    \midrule
    \texttt{Auto-play speed down}                    & Comma   \\
    \bottomrule
    \multicolumn{2}{p{.98\linewidth}}{\textbf{Note:} The ``Repeat current focal point'' function enables users to replay the sonification of the current focal point, the text, or both. The ``Auto-play'' function is typically used for sonification or braille, helping users quickly navigate through visualizations.}
  \end{tabular}
\end{table}

\section{Screen Reader Set Up}
\label{sec:appendix_presetup}
Participants will be guided to download the dictionary (refer to supplementary files) accustomed to different screen reader and browser, then they will start the setting up
\label{pre_settings}
\subsection{JAWS}
\begin{enumerate}
  \item Press number \texttt{2} to go to the JAWS session.
  \item Choose the preferred browser.
  \item Click the corresponding link.
  \item \texttt{control + J} open the download tab.
  \item Open the file folder.
  \item Cut the downloaded file.
        \begin{enumerate}
          \item Laptop Layout users press \texttt{control+shift+capslock+J} and Desktop users press \texttt{insert +J} to open JAWS interface
          \item Press \texttt{u} go to Utility
          \item Press \texttt{x} direct to Explore Utilities Folder
          \item Press down arrow key then click explore my settings and Paste the file.
        \end{enumerate}
  \item \texttt{alt+4} close the window.
  \item \texttt{control+W} close the Tab.
  \item \texttt{INSERT + V} open Quick Settings dialog
        \begin{enumerate}
          \item Type \texttt{Flash} and Uncheck the flash messages.
          \item \texttt{Control+E}.
          \item Type \texttt{eight} to Check the eight dot braille.
        \end{enumerate}
  \item Restart the JAWS.
  \item \texttt{Capslock+D} to check if the dictionary works.
\end{enumerate}

\subsection{NVDA}
\begin{enumerate}
  \item \texttt{control + J} open the download tab.
  \item Open the file folder.
  \item Cut the downloaded file.
  \item \texttt{Win+R}.
  \item Type \texttt{\%APPDATA\%} and hit enter.
  \item Go to nvda folder.
  \item Arrow down to folder \texttt{speechDicts}.
  \item Go to folder \texttt{appDicts}.
  \item Turn on Virtual Cursor.
        \begin{enumerate}
          \item \texttt{Control+Capslock+G}.
          \item \texttt{Shift+Tab}.
          \item braille $\rightarrow$ tab several times to find $\rightarrow$ show cursor $\rightarrow$ check $\rightarrow$ blink cursor $\rightarrow$ check.
        \end{enumerate}
  \item Restart the NVDA.
\end{enumerate}

\section{Tutorial}
\label{sec:appendix_tutorial}
In this section, we provide complete tutorial for bar plots, while for heatmaps, box plots, and scatter plots, we only include the unique part.

\subsection{Tutorial: Bar plot}
\subsubsection*{Navigation}

By default, all the modes are disabled, you can toggle on and off each mode by pressing \texttt{B} for braille, \texttt{T} for text, and \texttt{S} for sonification.

\begin{itemize}
  \item You can use the left and right arrow keys to navigate from bar to bar.
  \item You can use the combination of \texttt{Control + Home} and \texttt{Control + End} keys to quickly access the first bar and the last bar respectively.
\end{itemize}

\subsubsection*{Text}

Please press the \texttt{T} key to activate the text description (first press activate the terse mode, second press enables the verbose mode, and third press will deactivate text description). Depending on which mode you choose, terse or verbose, you will get different text descriptions.

\subsubsection*{Braille}

Please press the \texttt{B} key to activate (second press will deactivate) braille and now you can use the braille display to feel the plot.

\subsubsection*{Sonification}

\begin{itemize}
  \item You can press the \texttt{S} key to activate (the second press will deactivate) the sonification. The current system represents variations in value with pitch. The pitch raises when the y value increases and lowers when the y value decreases. The x-axis is represented in stereo panning sound from left to right and right to left, going along with x values.
  \item You can press the \texttt{Space} key to hear the sound of the current element again.
  \item You can also automatically play the sound for every bar. If you want to auto-play outward in the direction of the arrow, use \texttt{Control + Shift + Arrow} key.  If you want to auto-play inward in the direction of the arrow, use \texttt{Alt + Shift + Arrow key}.
  \item During the autoplay, you can use the \texttt{Control} key to stop the autoplay. You can also use the period key to speed up the autoplay or use the comma key to slow down the autoplay.
\end{itemize}

\subsubsection*{Help}

Whenever you forget the key combinations, you can always open the help menu by pressing the \texttt{H} key and close the help menu with the \texttt{Esc} key. Within the help menu, you can always change the settings including the volume, braille display size, autoplay rate, min frequency, and max frequency.

\subsection{Tutorial: Heatmap}

\subsubsection*{Navigation}

\begin{itemize}
  \item You can use the combination of \texttt{"Control"} key plus \texttt{"Arrow"} keys to quickly access the very far left/right/up/down cell.
  \item You can use the combination of \texttt{"Control + Home”/”End”} keys to quickly access the first cell (top left corner) or the last cell (bottom right corner).
\end{itemize}

\subsubsection*{Braille}

The braille will display all cells in an order from up left to bottom right, and the characters are separated by a row separator denoted with 1,2,5,6 braille chord.

\subsubsection*{Sonification}

You will hear an empty cell sound if the value in the cell equals to zero.

\subsection{Tutorial: Box plot}

\subsubsection*{Navigation}
The left/right arrow keys can be used to navigate within a boxplot. Using the up/down keys allows you to transition from one boxplot to another while maintaining focus on the same statistical concept. For example, if you are at the 25\% quantile of boxplot 1 and switch to boxplot 2, you'll be positioned at the 25\% quantile of boxplot 2. To quickly access the extreme left/right/up/down element, combine the \texttt{Control} key with the \texttt{Arrow} keys.

\subsubsection*{Braille}
In statistics, the range between the 25\% (Q1) and 75\% (Q3) points is termed the interquartile range. In boxplots, this is represented as a boxed area. In our system, this box area is depicted as a rectangular shape characterized by the dots 1, 2, 3, 4, 5, 6 in the braille chord pattern. The broader this box area, the more of these rectangular patterns will be discerned. A vertical line divides the interquartile range box, representing the 50\% data point or median (Q2). This line is symbolized in braille with the chords dots 4, 5, 6 and dots 1, 2, 3. The whisker line on the left visually represents the range from the 25\% point to the minimum value and is signified with the dots 2, 5 braille chord. A wider whisker line means a more frequent appearance of repeated line patterns. Similarly, the range from the 75\% point to the maximum value, represented visually by a whisker line on the right, is also denoted with the dots 2, 5 braille chord. Data points outside the minimum and maximum values are outliers, and each outlier point is symbolized with dot 2. The more outliers present in the data, the more one will perceive a scattered-dot pattern.

\subsubsection*{Sonification}
When navigating to outliers, users will encounter an auto-play sound representing all outliers.

\subsection{Tutorial: Scatter plot}
\subsubsection*{Layer Switch}
You can use “\texttt{PageUp}” and “\texttt{PageDown}” keys to switch different layers of the scatter plot.

\subsubsection*{Layer 1: Sonification}
You can press the \texttt{S} key to activate the sonification. The first press activates separate sonification, the second enables combined sonification, and the third will deactivate the sonification.

\subsubsection*{Layer 2: Sonification without braille enabled}
You can only use “\texttt{Control + Shift + Arrow}” key or “\texttt{Alt + Shift + Arrow}” key to hear the bending sound.

\subsubsection*{Layer 2: Braille}
Please press the \texttt{B} key to activate (the second press will deactivate) braille. You can then use the braille display to feel the trend line.

\subsubsection*{Layer 2: Sonification with braille enabled}
After activating the braille, you can proceed with the following. To automatically play the sound for every point along the smoothed line, use “\texttt{Control + Shift + Arrow}” key for outward auto-play in the direction of the arrow, or “\texttt{Alt + Shift + Arrow}” key for inward auto-play.

\section{Task Questions}
\label{sec:appendix_taskql}
\subsubsection*{Bar plot}
\begin{enumerate}
  \item Which continent has the largest population and its corresponding value?
  \item Which continent has the lowest population and its corresponding value?
  \item Between Africa and Americas, which has the higher population?
  \item Between Africa and Europe, which has the lower population?
  \item Can you describe the shape of the bar plot?
\end{enumerate}

\subsubsection*{Heatmap}
\begin{enumerate}
  \item Which continent in which year has the highest average GDP and its corresponding value?
  \item Where do you find the highest average GDP? You can use the “left” “center” “right” combined with “top” “middle” “bottom” to describe.
  \item Which continent in which year has the lowest average GDP and its corresponding value?
  \item Where do you find the lowest average GDP? You can use the “left” “center” “right” combined with “top” “middle” “bottom” to describe.
  \item How do you interpret the cell row =2, column=3?
  \item Can you describe any patterns in the heatmap?
\end{enumerate}

\subsubsection*{Box plot}
\begin{enumerate}
  \item By comparing Oceania and Europe, which continent has the higher 25\% value of life expectancy?
  \item By comparing Oceania and Africa, which continent has the lower 50\% value of life expectancy?
  \item Which continent has the most number of lower outliers?
  \item Regarding Asia, is the data of life expectancy skewed?
\end{enumerate}

\subsubsection*{Scatter plot}
\begin{enumerate}
  \item Do GDP and Life Expectancy have a positive correlation or negative correlation?
  \item Do GDP and Life Expectancy have a strong, moderate, or weak relationship?
  \item Can you elaborate on the shape of the trend line?
  \item Can you describe the distribution of the points?
\end{enumerate}

\section{State Diagrams}
\label{sec:appendix_statediagram}
Using individual interaction patterns as a foundation, we created aggregated state diagrams. In these state diagrams, ellipses represent states and edges illustrate the transitions between them. The breadth of an edge and its corresponding label indicate the number of participants following that specific transition, while unique colors differentiate each task question. 
\newpage
\begin{figure*}[!t]
  \centering
  \includegraphics[width=1\linewidth]{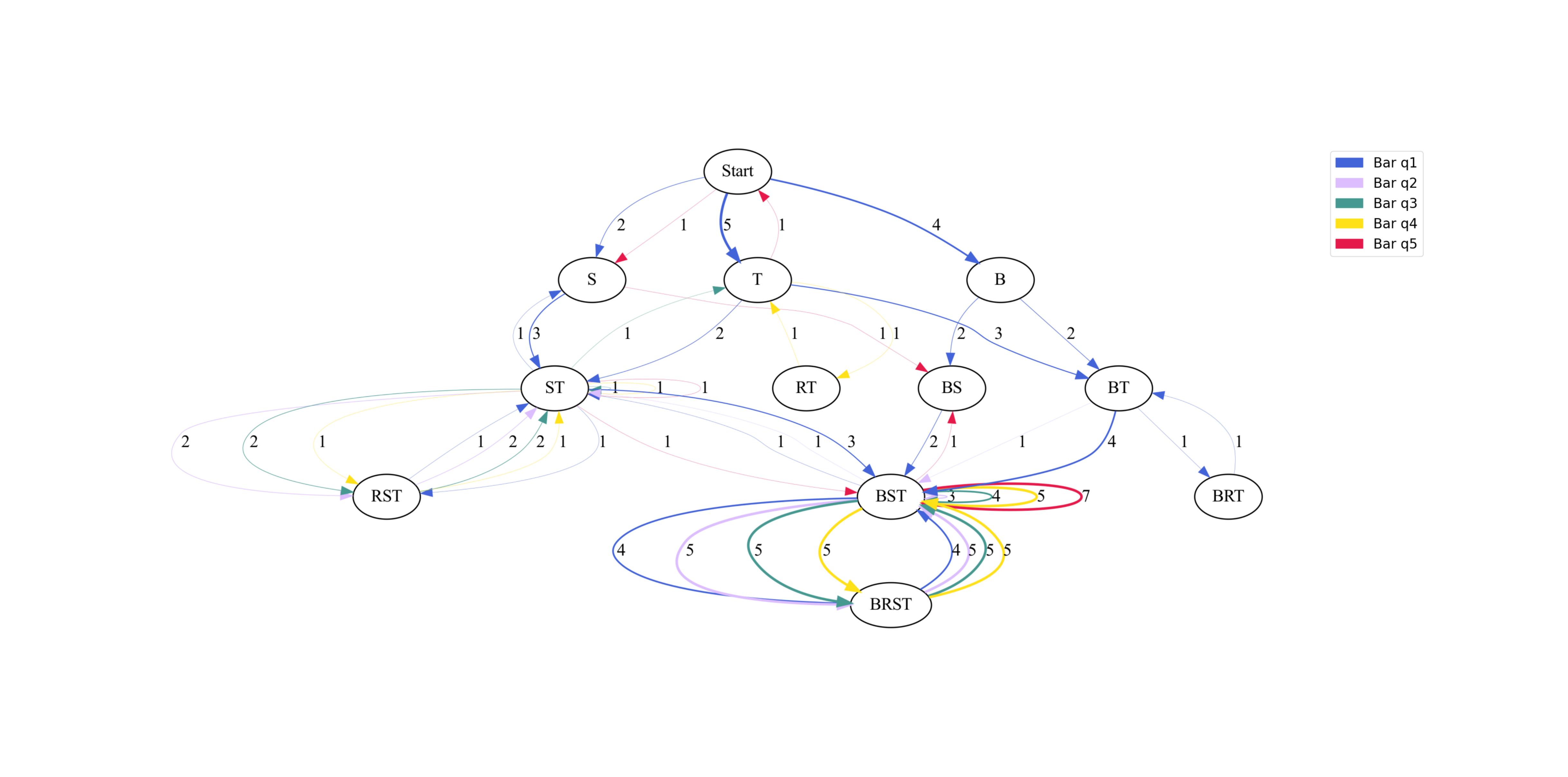}
  \caption{State Diagram for Participate Interactions with Bar plot during the Task Session.}
  \Description{
  This image is a state diagram that represents the different modes supported by MAIDR. The modes are indicated by nodes labeled with combinations of the letters B, T, S, and R, which stand for Braille, Text, Sonification, and Review respectively. The diagram begins at a node labeled "Start," and from there, transitions to other nodes based on the modes being represented. For example, the node labeled "ST" would represent a state where both Sonification and Text are active. Arrows with numbers point from one node to another, indicating the possible transitions between states. The numbers on the arrows represent the number of participants  who have the same interaction i.e., change from one state to another.
  
  The diagram also includes colored bars with the labels "bar q1" through "bar q5" in a legend to the right, which likely correspond to different types of questions for bar plot during the interview. 
  }
  \label{fig:state_diagram_bar}
\end{figure*}

\begin{figure*}[ht]
  \centering
  \includegraphics[width=1\linewidth]{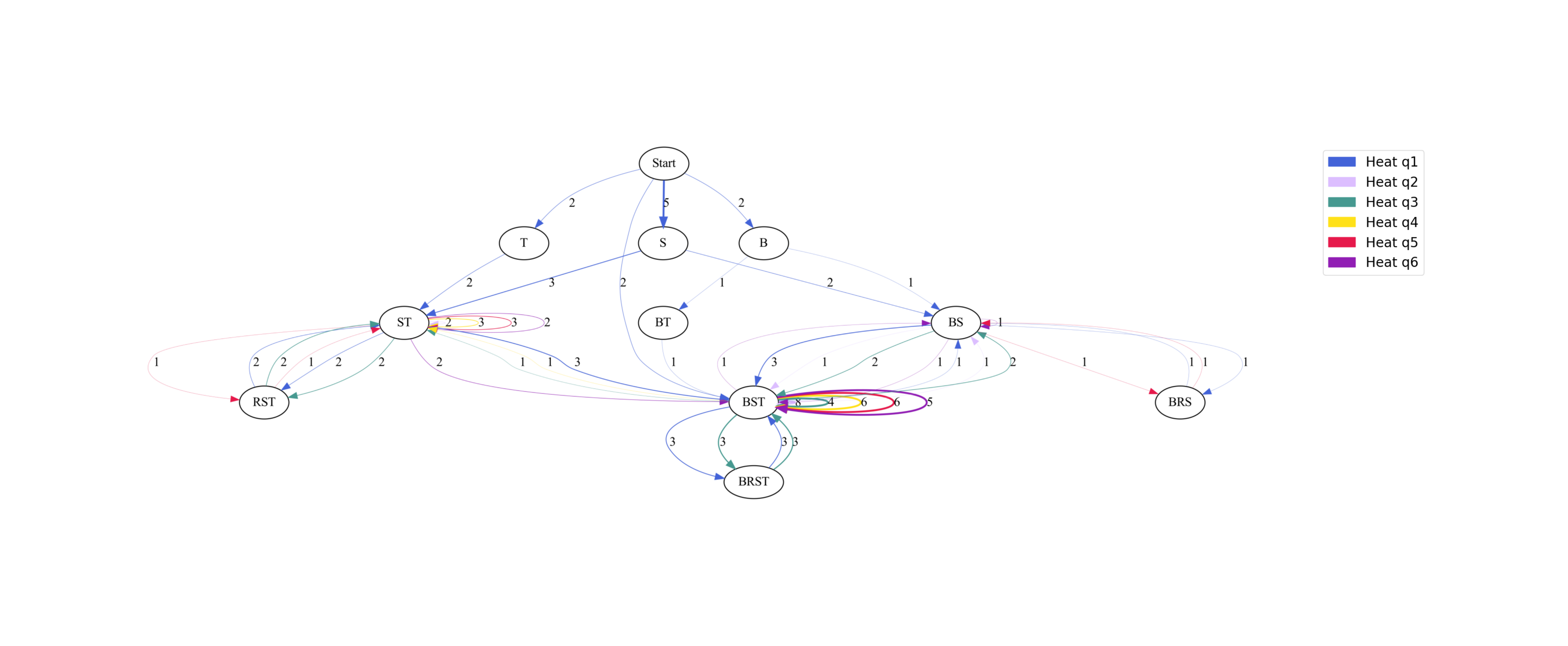}
  \caption{State Diagram for Participate Interactions with heatmap during the Task Session.}
  \Description{
  This image is a state diagram that represents the different modes supported by MAIDR. The modes are indicated by nodes labeled with combinations of the letters B, T, S, and R, which stand for Braille, Text, Sonification, and Review respectively. The diagram begins at a node labeled "Start," and from there, transitions to other nodes based on the modes being represented. For example, the node labeled "ST" would represent a state where both Sonification and Text are active. Arrows with numbers point from one node to another, indicating the possible transitions between states. The numbers on the arrows represent the number of participants  who have the same interaction i.e., change from one state to another.
  
  The diagram also includes colored bars with the labels "heat q1" through "heat q6" in a legend to the right, which likely correspond to different types of questions for heat map during the interview.
  }
  \label{fig:state_diagram_heat}
\end{figure*}

\begin{figure*}[ht]
  \centering
  \includegraphics[width=1\linewidth]{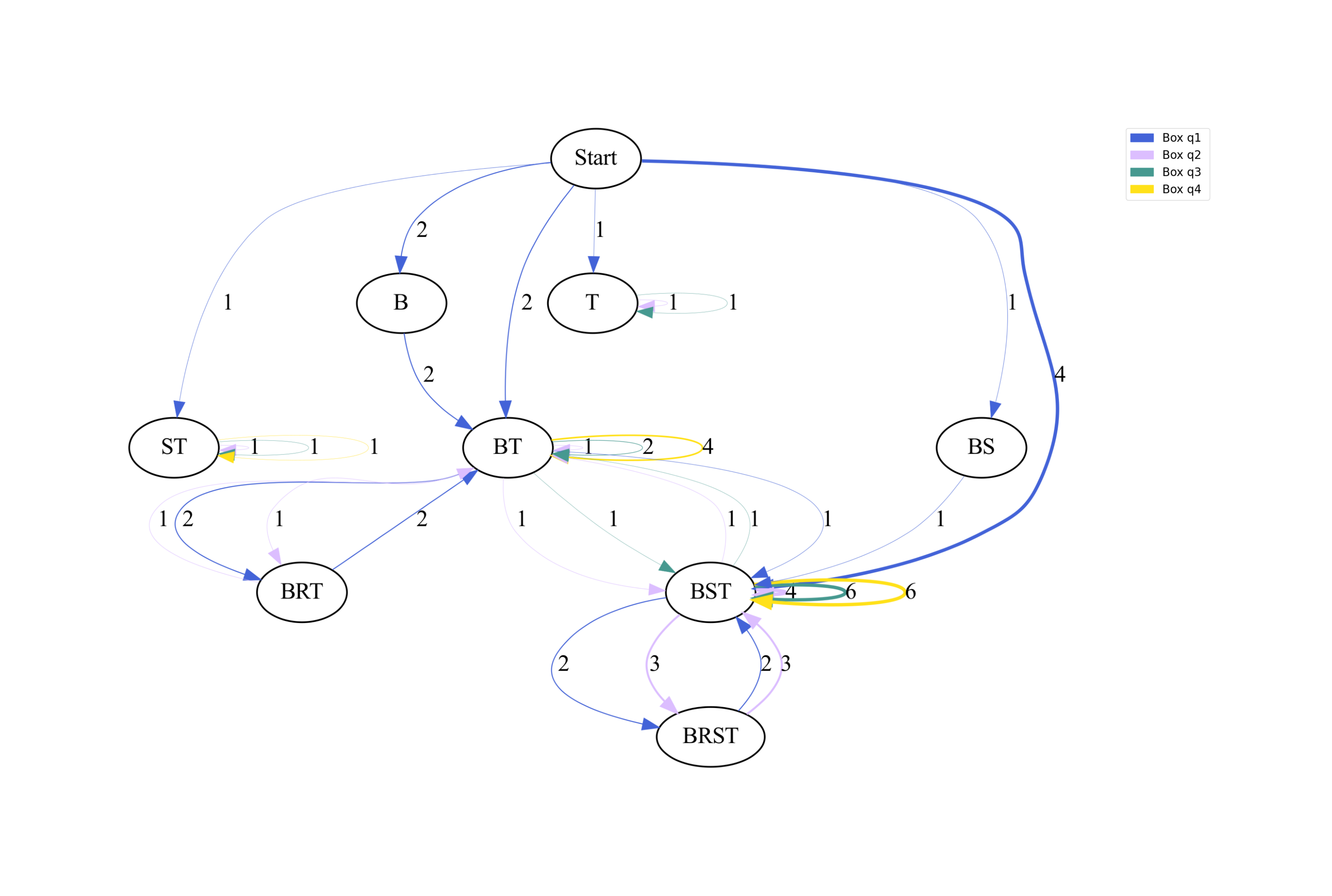}
  \caption{State Diagram for Participate Interactions with Box plot during the Task Session.}
  \Description{
  This image is a state diagram that represents the different modes supported by MAIDR. The modes are indicated by nodes labeled with combinations of the letters B, T, S, and R, which stand for Braille, Text, Sonification, and Review respectively. The diagram begins at a node labeled "Start," and from there, transitions to other nodes based on the modes being represented. For example, the node labeled "ST" would represent a state where both Sonification and Text are active. Arrows with numbers point from one node to another, indicating the possible transitions between states. The numbers on the arrows represent the number of participants  who have the same interaction i.e., change from one state to another.
  
  The diagram also includes colored bars with the labels "box q1" through "box q4" in a legend to the right, which likely correspond to different types of questions for box plot during the interview.
  }
  \label{fig:state_diagram_box}
\end{figure*}

\begin{figure*}[ht]
  \centering
  \includegraphics[width=1\linewidth]{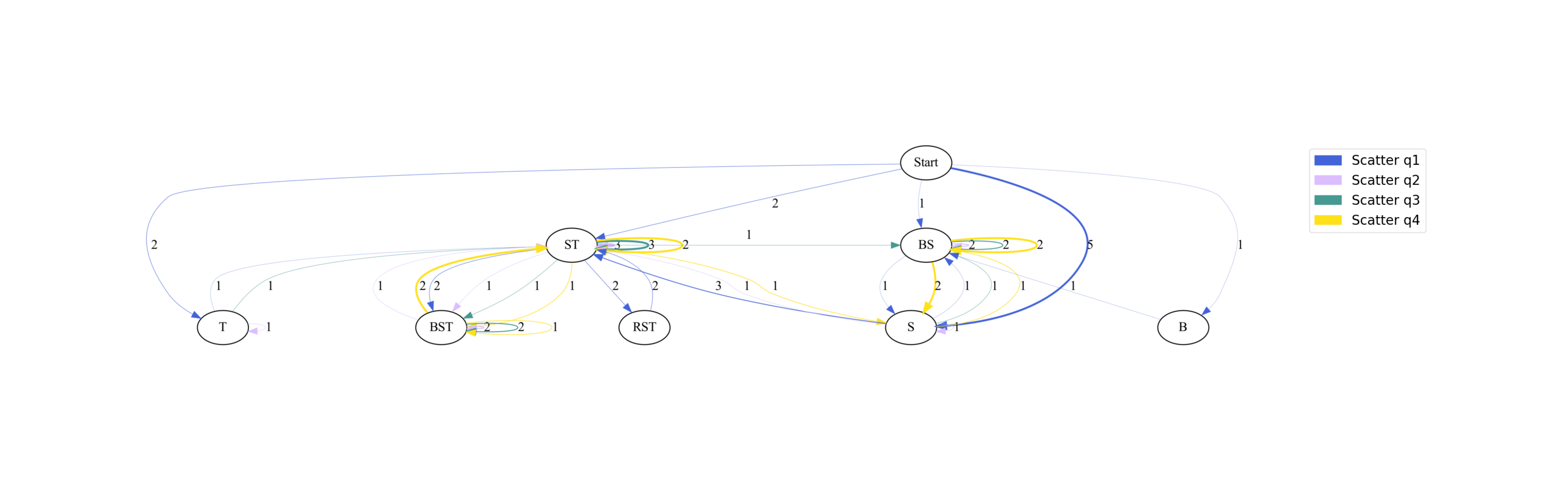}
  \caption{State Diagram for Participate Interactions with Scatter plot during the Task Session.}
  \Description{
  This image is a state diagram that represents the different modes supported by MAIDR. The modes are indicated by nodes labeled with combinations of the letters B, T, S, and R, which stand for Braille, Text, Sonification, and Review respectively. The diagram begins at a node labeled "Start," and from there, transitions to other nodes based on the modes being represented. For example, the node labeled "ST" would represent a state where both Sonification and Text are active. Arrows with numbers point from one node to another, indicating the possible transitions between states. The numbers on the arrows represent the number of participants who have the same interaction i.e., change from one state to another.
  
  The diagram also includes colored bars with the labels "scatter q1" through "scatter q4" in a legend to the right, which likely correspond to different types of questions for scatter plot during the interview.
  }
  \label{fig:state_diagram_scatter}
\end{figure*}